\documentclass{iopart}
\pdfoutput=1
\usepackage{euscript,iopams,amssymb,amsfonts,graphicx,bm}
\usepackage{setstack}

  \usepackage{epstopdf}
  \usepackage{graphics}

\usepackage{color}

\usepackage{esint}
\eqnobysec

\newcommand{\E}{{\mathbb E}}

\newcommand{\z}{\mathbf{z}}

\renewcommand{\r}{\mathbf{r}}
\renewcommand{\u}{\mathbf{u}}
\renewcommand{\v}{\mathbf{v}}
\newcommand{\U}{\mathbf{U}}
\newcommand{\calJ}{{\mathcal J}}
\newcommand{\calN}{{\mathcal N}}
\newcommand{\calP}{{\mathcal P}}
\newcommand{\calU}{{\mathcal U}}

\newcommand{\calH}{\mathbf{H}}

\newcommand{\q}{\mathbf{q}}

\newcommand{\Markov}[2]{\underset{#1}{\overset{#2}{\rightleftharpoons}}}
\newcommand{\x}{\mathbf{x}}

\renewcommand{\e}{{\rm e}}
\def\Z{{\mathbb Z}} 
 
\def\P{{\mathbb P}} 
 
\def\R{{\mathbb R}}

\renewcommand{\theequation}{\arabic{section}.\arabic{equation}}

\begin{document}

\title[Stochastic hybrid path integrals]{Path integrals for stochastic hybrid reaction-diffusion processes} 
\author{Paul C. Bressloff$^{1}$ }
\address{$^1$Department of Mathematics, University of Utah, 155 South 1400 East, Salt Lake City, Utah 84112, USA}
\ead{bressloff@math.utah.edu}

\date{\today}

\begin{abstract} We construct path integrals for stochastic hybrid reaction-diffusion (RD) processes, in which the reaction terms depend on the discrete state of a randomly switching environment. We proceed by spatially discretizing a given RD system and using a spinor representation of the environmental states to derive a path integral for the lattice model. In the case of large molecular numbers, the corresponding continuum path integral action is expressed in terms of an effective Hamiltonian, which involves a concentration field $u(\x,t)$, $\x\in \R^d$, a conjugate field $v(\x,t)$, and $M$ auxiliary conjugate pairs $(c_m(t),\phi_m(t))$, where $M$ is the number of discrete environmental states. The variable $c_m(t)$ determines the effective probability that a sample path is exposed to the $m$-th environmental state at time $t$, with $\sum_{m=1}^Mc_m(t)=1$. We then consider the semi-classical (adiabatic) limit $\epsilon \rightarrow 0$, where $\epsilon^{-1}$ determines the rate of switching between the environmental states. We show how the auxiliary variables can be eliminated to yield an action functional for the fields $u$ and $v$ alone. The associated Hamiltonian is the sum of a diffusion term and the Perron or principal eigenvalue of a functional linear operator involving the reaction terms and the matrix generator of the switching process. The reduced path integral is then used to derive a functional Hamilton-Jacobi equation for least action paths and to obtain a Gaussian noise approximation of the stochastic hybrid RD system in the adiabatic limit.
The theory is illustrated using a model of diffusion on a two-dimensional substrate that switches between an active and an inactive state. Finally, the path integral in the case of low molecular numbers is constructed by considering a corresponding RD master equation. It is now necessary to take into account two sources of noise, one due to the switching environment and the other due to fluctuations in molecular numbers. In particular, one has to specify the $\epsilon$-scaling of both sources in the semi-classical limit.

 \end{abstract}


\maketitle

\section{Introduction}\label{sec:Intro}

Diffusion processes in randomly switching environments are finding increasing applications in cell biology and biophysics. Examples include diffusion in domains with stochastically-gated boundaries \cite{Berez02,Berez03,Holcman10,Lawley15,Bressloff15a,Bressloff15c}, diffusion over fluctuating barriers \cite{Agmon84,Doering92,Reimann98,Ankerhold98}, and stochastic gap junctions \cite{Bressloff16}. Mathematically speaking, diffusion in a randomly switching environment is an infinite-dimensional version of a so-called stochastic hybrid system. Stochastic hybrid systems involve a coupling between a discrete Markov chain and a continuous stochastic process \cite{Bressloff14a}. If the latter evolves deterministically between jumps in the discrete state, then the system reduces to a piecewise deterministic Markov process (PDMP) \cite{Davis84}. Well known examples of finite-dimensional hybrid systems include stochastic gene expression \cite{Kepler01,Bose04,Newby12,Newby15,Hufton16,Bressloff17a}, voltage fluctuations in neurons \cite{Fox94,Chow96,Keener11,Goldwyn11,Buckwar11,NBK13,Bressloff14b,Newby14}, and motor-driven intracellular transport \cite{Reed90,Friedman05,Newby10,Bressloff13}. One method for analyzing the diffusion equation with switching boundary conditions is to discretize space and construct the Chapman-Kolmogorov (CK) equation for the resulting finite-dimensional stochastic hybrid system \cite{Bressloff15a}. The CK equation can then be used to derive moment equations for the stochastic concentration. In the continuum limit, this yields a hierarchy of moment equations, with the equation at $r$-th order taking the form of an $r$-dimensional parabolic partial differential equation (PDE) that couples to lower order moments at the boundaries. Although the diffusing particles are non-interacting, statistical correlations arise at the population level due to the fact that they all move in the same randomly switching environment. 

We have previously developed path integral methods for studying finite-dimensional stochastic hybrid systems in the weak noise (adiabatic) limit. In particular, we have shown how to derive a hybrid path integral using two alternative methods: (i) integral representations of the Dirac delta function \cite{Bressloff14,Bressloff15}, which is analogous to the construction of path integrals for stochastic differential equations (SDEs) \cite{Martin73,Dom76,Janssen76}; (ii) bra-kets and ``quantum-mechanical'' operators \cite{Bressloff21a}, similar in spirit to the Doi-Peliti formalism for master equations \cite{Doi76,Doi76a,Peliti85,Weber17}. In both cases the Hamiltonian of the resulting action functional corresponds to the principal eigenvalue of a linear operator, which combines the generator of the discrete Markov process and the vector fields of the piecewise deterministic dynamics. This is consistent with more rigorous results obtained using large deviation theory \cite{Kifer09,fagg09,Faggionato10,Bressloff17}.

Parallel to the development of operator and path integral methods for finite-dimensional hybrid systems, there has been a series of studies of stochastic gene expression in the presence of promoter noise and low protein copy numbers \cite{Sasai03,Zhang13,Chen15,Li16,Bhatt20}. In these examples, a single gene network typically consists of two discrete variables, one specifying the activity state of the gene and the other the number of proteins. Although the resulting system evolves according to a continuous time Markov chain, and is thus not strictly a hybrid system, there is a separation of time scales between the fast switching of the gene state and the relatively slow synthesis and degradation of the protein. (Note that by carrying out a system size expansion with respect to the number of proteins, one could reduce the dynamics of the protein concentration to an SDE and thus obtain a true stochastic hybrid system.) In the case of a stochastic gene network, one can construct an operator version of the corresponding chemical master equation by representing fluctuations in protein concentrations in terms of Doi-Peliti bosonic operators, and projecting the discrete activity state of the gene onto a coherent spin state \cite{Sasai03,Zhang13,Bhatt20} or a more general spinor representation \cite{Chen15,Li16}. (A coherent spin state is a particular type of spinor that is parameterized on the 2-sphere \cite{Radcliffe71,Fradkin13}.) The resulting path integral action involves auxiliary coordinate and momentum variables arising from the parameterization of the spin states. The analogy with quantum spin systems also allows variational methods to be used to approximate the energy landscape of the genetic switch \cite{Sasai03}, although care must be taken since the effective Hamiltonian operator of the master equation is non-Hermitian.

In this paper, we use the general spinor formalism developed for gene networks to construct path integrals for stochastic hybrid reaction-diffusion (RD) processes, under the assumption that the switching environment affects the reaction term rather than the boundary conditions. We begin in section 2 by defining a piecewise deterministic partial differential equation (PDE) for a stochastic hybrid RD process, and deriving moment equations for the concentration. In section 3 we present a detailed derivation of the path integral for the spatially discretized hybrid RD model. Taking the continuum limit then yields a functional path integral for the original stochastic hybrid RD equation. The associated action functional can be expressed in terms of an effective Hamiltonian involving the concentration field $u(\x,t)$, $\x\in \R^d$, its conjugate field $v(\x,t)$, and $M$ auxiliary conjugate pairs $(c_m(t),\phi_m(t))$, where $M$ is the number of discrete environmental states. The variable $c_m(t)$ determines the effective probability that a sample path is exposed to the $m$-th environmental state at time $t$ with $\sum_{m=1}^Mc_m(t)=1$.  In section 4, we consider the semi-classical (adiabatic) limit $\epsilon \rightarrow 0$, where $\epsilon^{-1}$ determines the rate of switching between the environmental states. We show how the auxiliary variables can be eliminated to yield an action functional for the fields $u$ and $v$ alone. The associated Hamiltonian is of the form $H[u,v]=D\int_{\R^d} v(\x) \nabla^2 u(\x)d\x+\Lambda[u,v]$ where $\Lambda$ is the Perron or principal eigenvalue of a functional linear operator involving the reaction terms and the matrix generator of the switching process. The reduced path integral is then used to derive a functional Hamilton-Jacobi equation for least action paths and to obtain a Gaussian noise approximation of the stochastic hybrid RD system in the adiabatic limit. In section 5 we illustrate the theory by considering diffusion over a two-dimensional substrate that switches between an active and an inactive state. Finally, in section 6 we extend our analysis to a hybrid RD master equation on a lattice, in which diffusive hopping between neighboring lattice sites is treated as an additional set of single step reactions that supplement the local chemical reactions. This type of model is necessary when the number of molecules at each lattice site is relatively small. We construct the corresponding hybrid path integral along analogous lines to the hybrid PDE model, and derive the effective path integral action in the semi-classical limit. However, it is now necessary to take into account two sources of noise, one due to the switching environment and the other due to fluctuations in molecular numbers. In particular, we need to specify the $\epsilon$-scaling of both sources of noise in the semi-classical limit.

\section{Stochastic hybrid PDE}

Consider a one-dimensional RD process in which the reaction term depends on the current discrete state of the environment, which is denoted by $N(t)\in \{1,\ldots,M\}$. The latter is assumed to evolve according to an M-state irreducible Markov chain with matrix generator ${\bf Q}$. \footnote{A discrete process is said to be irreducible if there exists a $t>0$ such that $\e^{{\bf Q}t}>0$; this implies that any two states of the Markov chain can be connected in a finite time. One can then apply the Perron-Frobenius theorem for finite square matrices. In particular, there exists a unique positive right-eigenvector $\rho_n$ for which $\sum_{m}Q_{nm}\rho_m=0$; the corresponding left eigenvector is $(1,1,\ldots,1)$ since $\sum_{n}Q_{nm}=0$. We can identify $\rho$ as the unique stationary density. Moreover, the Perron Frobenius theorem ensures that all other eigenvalues have negative real parts, ensuring that the distribution $P_m(t)\rightarrow \rho_m$ as $t\rightarrow \infty$.}. Note that
\begin{equation}
\sum_{n=1}^{M}Q_{nm} =0,\quad \sum_{n=1}^{M}Q_{nm}(x)\rho_m =0
\end{equation}
for all $x\in \R$, where ${\mathbf \rho} $ is the stationary distribution of the Markov chain. The generator ${\bf Q}$ can be expressed in  terms of the corresponding transition matrix ${\bf T}$ according to
\begin{equation}
\label{QW}
Q_{nm}=T_{nm}-\delta_{n,m}\sum_{l=1}^{M}T_{ln},\quad T_{nn}=0.
\end{equation}
Setting $P_n(t)=\P[N(t)=n]$, we have the master equation
\begin{equation}
\label{master}
\frac{dP_n}{dt}=\sum_{m=1}^MQ_{nm}P_m(t).
\end{equation}
In between jumps in the environmental state, with $N(t)=n$, the concentration $U(x,t)$ evolves according to the piecewise RD equation 
\begin{eqnarray}
\label{swdiff}
  \frac{\partial U}{\partial t}&=D\frac{\partial^2U}{\partial x^2}+F_n(U),\quad x\in \Omega \subseteq \R.
\end{eqnarray}
Note that $U(x,t)$ is a stochastic concentration field. In the case of a finite interval, $\Omega=[0,L]$, the PDE has to be supplemented by boundary conditions at $x=0, L$. In previous work \cite{Bressloff15a}, we took the boundary conditions themselves to depend on the environmental state (with $F_n=0$). That is,
\numparts
\begin{eqnarray}
 b_nU(0,t)+c_n \partial_xU(0,t)&=d_n,\\ b'_nU(L,t)+c'_n \partial_xU(L,t)&=d'_n
 \end{eqnarray}
 \endnumparts
for $N(t)=n$ and constant coefficients $b_n$, $b_n'$, $c_n$, $c_n'$, $d_n$, $d_n'$. We showed that one way to analyze the effects of a random environment is to spatially discretize the PDE so that it is converted to a stochastic hybrid ODE \cite{Bressloff15a}. Introducing the lattice spacing $h$ and
setting $u_j=u(jh)$, $j\in \calJ =\{ 0, 1,\ldots,  \cal{N}\}$ with ${\cal N}a=L$, leads to the piecewise deterministic ODE
\begin{equation}
\label{sh}
\frac{dU_i}{dt}=\sum_{j\in \calJ}\Delta_{ij}^nU_j +F_n(U_i),\quad i\in \calJ,
\end{equation}
where $\U=(U_j,\, j \in \calJ)$ and $\Delta^n_{ij}$ is the discrete Laplacian for $N(t)=n$. Away from the boundaries, $\Delta^n_{ij}=\Delta_{ij}$ with
\begin{equation}
\label{dL1}
\Delta_{ij}=\frac{D}{h^2}[\delta_{i,j+1}+\delta_{i,j-1}-2\delta_{i,j}],\quad 0<i<\calN,
\end{equation}
whereas $\Delta^n_{ij}$ is modified at the boundaries $i=0,\calN$ in order to be consistent with the boundary conditions (see \cite{Bressloff15a} for details).
The corresponding probability density
\begin{equation}
 \mbox{Prob}\{\U(t)\in (\u,\u+d\u), N(t)=n\}={\mathcal P}_n(\u,t)d\u
\end{equation}
 evolves according to a differential Chapman-Kolmogorov (CK) equation for the stochastic hybrid system (\ref{sh}):
\begin{eqnarray}
\label{swCK0}
\fl  \frac{\partial \calP_n}{\partial t}&=-\sum_{i\in \calJ}\frac{\partial}{\partial u_i}\left [\left (\sum_{j\in \calJ}\Delta^n_{ij}u_j+F_n(u_i)\right )\calP_n(\u,t)\right ] +\sum_{m=1}^MQ_{nm}\calP_m(\u,t).
\end{eqnarray}

Now define the conditional first moments
\begin{equation}
{\mathcal U}_{n,k}(t)=\E[U_k(t)1_{N(t)=n}]=\int d\u \,\calP_n(\u,t)u_k(t),
\end{equation}
where $\int d\u=\prod_j \int_{-\infty}^{\infty}du_j $.
Multiplying both sides of the CK equation (\ref{swCK0}) by $u_k(t)$ and integrating by parts with respect to $\u$ yields the moment equation
\begin{equation}
\label{mom1d}
\frac{d {\mathcal U}_{n,k}}{d t}=\sum_{j=1}^{\calN}\Delta^n_{kj}{\mathcal U}_{n,j}+\E[F_n(U_k)1_{N(t)=n}] + \sum_{m=1}^MQ_{nm}{\mathcal U}_{m,k}.
\end{equation}
For the sake of illustration, suppose that $F_n(u)=-\gamma_n u$ where $\gamma_n$ is some environmentally-dependent adsorption rate \cite{Bressloff17}, see also section 5. In this case, $\E[F_n(U_k)1_{N(t)=n}]=-\gamma_n {\mathcal U}_{n,k}$, and Eq. (\ref{mom1d}) becomes a closed equation for the first moments. Retaking the continuum limit $h\rightarrow 0$ then yields the coupled system of deterministic PDEs for ${\mathcal U}_n(x,t)=\E[U(x,t) 1_{N(t)=n}]$
\begin{eqnarray}
\label{mom1}
   \frac{\partial {\mathcal U}_n}{\partial t}&=D\frac{\partial^2{\mathcal U}_n(x,t)}{\partial x^2}-\gamma_n {\mathcal U}_n(x,t)+ \sum_{m=1}^MQ_{nm}{\mathcal U}_{m}(x,t)
\end{eqnarray}
for $x\in [0,L]$, with the boundary conditions
\begin{eqnarray}
 \fl  b_n\calU_n(0,t)+c_n \partial_x\calU_n(0,t)&=d_n,\quad b'_n\calU_n(L,t)+c'_n \partial_x\calU_n(L,t)=d'_n.
 \end{eqnarray}
As we have previously highlighted \cite{Bressloff15a}, since all the particles diffuse in the same randomly switching environment, there are non-trivial statistical correlations between the particles. For example, consider the second-order moments
\begin{equation}
C_n(x,y,t)=\E[U(x,t)U(y,t)1_{N(t)=n}].
\end{equation}
These evolve according to the moment equation
  \begin{eqnarray}
  \label{C22}
  \frac{\partial C_n}{\partial t} &=D\frac{\partial^2 C_n}{\partial x^2}+D\frac{\partial^2C_n}{\partial y^2 }-2\gamma_nC_n+\sum_{m\in I}Q_{nm}C_m,
  \end{eqnarray}
  together with boundary conditions that couple to the first-order moments \cite{Bressloff15a}.
The latter can be derived from the spatially discretized CK equation (\ref{swCK0}) after multiplying both sides by the product $u_k(t)u_l(t)$, integrating by parts and retaking the continuum limit. Clearly $C_n(x,y,t)\neq \calU_n(x,t)\calU_n(y,t)$, which means that the two-point correlation function is non-zero. A similar comment holds for higher-order moments.
\medskip

\setcounter{equation}{0}
\section{Construction of a spatially discretized hybrid path integral}

In this section we construct a path integral representation for the spatially discretized hybrid system (\ref{sh}). We proceed by adapting the spinor representation introduced in Refs. \cite{Chen15,Li16} within the context of gene networks. For convenience, we ignore boundary effects by taking $x\in \R$ in equation (\ref{swdiff}) so that the lattice in equation (\ref{sh}) is $\calJ=\Z$. Alternatively, we could take $x\in [0,L]$ and impose periodic boundary conditions so that $\calJ$ is finite. In both cases, the discrete Laplacian is given by equation (\ref{dL1}) for all $i\in \calJ$ and is thus independent of the environmental state. 

\subsection{Spinor representation}

Consider the master equation (\ref{master}) for an $M$-state Markov chain, written in matrix form
\begin{equation}
\label{mc}
\frac{d{\bf P}}{dt}={\bf Q}{\bf P}(t),\quad {\bf P}(t)=(P_1(t),P_2(t)),\ldots,P_M(t)^{\top}.
\end{equation}
Introduce the multicomponent spinors
\begin{equation}
\label{spinor}
|s\rangle=\left (\begin{array}{c} c_1\e^{i\phi_1/2}  \\ \vdots \\ c_M\e^{i\phi_M/2} \end{array} \right ),\quad \langle s| =\left (\begin{array}{ccc} \e^{-i\phi_1/2} &\ldots & \e^{-i\phi_M/2}  \end{array} \right ),
\end{equation}
with $c_j\in [0,1]$, $\phi_j\in [0,2\pi)$  and the normalization condition
\begin{equation}
\sum_{m=1}^{M}c_m=1.
\end{equation}
(In addition, we are free to set $\sum_{m=1}^M\phi_m=0
$.)
Note that 
\begin{equation}
\langle s'|s\rangle =\sum_{m=1}^{M}\e^{i(\phi_m-\phi'_m)/2}c_m,
\end{equation}
so that $\langle s|s\rangle =1$ and
\begin{equation}
\label{sdiff}
\langle s+\Delta s |s\rangle =1-\frac{1}{2}i \sum_{m=1}^{M}c_m\Delta \phi_m+O(\Delta \phi^2).
\end{equation}
We also have the completeness relation
\begin{equation}
\label{comp2}
\prod_{m=1}^{M-1}\left \{\int_0^{1} \frac{dc_m }{2}\, \int_0^{4\pi}\frac{d\phi_m}{4\pi}\right \} |s\rangle \langle s|=1.
\end{equation}
In terms of the transition elements $T_{nm}$, we can write
\begin{eqnarray}
\langle s|{\bf Q}|s\rangle =Q({\bf c},{\bm \phi}):= \sum_{\underset{n\neq m}{n,m=1}}^MT_{nm}\left [\e^{i(\phi_m-\phi_n)/2}-1\right ]c_m.
\end{eqnarray}

In the special case of a 2-state hybrid system ($M=2$), we can set $c_1=z,c_2=1-z$ and $\phi_1 =\phi=-\phi_2$. Under the change of variables $z=\cos ^2\theta /2$, $0\leq \theta \leq \pi$, the spinor can be rewritten as
\begin{equation}
|s\rangle=\left (\begin{array}{c} \e^{i\phi /2} \cos ^2\theta /2 \\  \e^{-i\phi/2}\sin ^2\theta /2 \end{array} \right ),\quad \langle s| =\left (\begin{array}{ccc} \e^{-i\phi }   & \e^{i\phi }  \end{array} \right ),
\end{equation}
which is a representation of the coherent spin-1/2 state on the unit sphere \cite{Radcliffe71,Sasai03,Zhang13,Bhatt20}. The transition matrix has the general form
\begin{equation}
{\bf T}=\left (\begin{array}{cc} 0 & \alpha \\ \beta &0 \end{array}\right ), \quad \alpha,\beta \geq 0,
\end{equation}
so that
\begin{eqnarray}
\langle s|{\bf Q}|s\rangle =-\beta \left (1-\e^{i\phi}\right )\frac{1+\cos\theta}{2} -\alpha \left (1-\e^{-i\phi}\right )\frac{1-\cos\theta}{2} .
\end{eqnarray}
On the other hand, if $M>2$ then the corresponding coherent spin-$S$ states ($M=2S+1$) form a restricted class of $M$-component spinors. For example, suppose that $M=3$. Restricting the probability variables $z_m$ to have the parametric form 
\begin{equation}
c_1=z^2,\quad c_2=2z(1-z),\quad c_3 = (1-z)^2,
\end{equation}
with $z=\cos^2\theta/2$, and setting $\phi_1=2\phi, \phi_2=0,\phi_3=-2\phi$, yields the coherent spin-1 state
\begin{equation}
|s\rangle = \left (\begin{array}{c} \e^{i\phi}\cos^4\theta/2  \\ 2\cos^2\theta/2\, \sin^2\theta/2 \\ \e^{-i\phi}\sin^4\theta/2\end{array} \right ),\quad 0\leq \theta \leq \pi,\ 0\leq \phi <2\pi,
\end{equation}
together with the adjoint $\langle s|=\left (  \e^{-i\phi}, \, 1,\, \e^{i\phi} \right ).$ 
Given the general transition matrix
\begin{equation}
\label{Q3}
{\bf T}=\left (\begin{array}{ccc}0 &\alpha_- &\gamma_+  \\ \alpha_+ &0& \beta_- \\ \gamma_- & \beta_+ &0 \end{array} \right ),
\end{equation}
we have
\begin{eqnarray}
\label{Qop3}
 \langle s|{\mathbf Q}|s\rangle &=   - \left (1-\e^{i\phi}\right )\left ( 2\beta_+\cos^2\theta/2\, \sin^2\theta/2+\alpha_+\cos^4\theta/2\right )\nonumber \\
  &\quad  -\left (1-\e^{-i\phi}\right )\left ( 2\alpha_-\cos^2\theta/2\, \sin^2\theta/2+\beta_-\sin^4\theta/2\right )
\nonumber
\\
  &\quad  -\gamma_+\left (1-\e^{-2i\phi}\right )\sin^2\theta/2-\gamma_-\left (1-\e^{2i\phi}\right )\cos^2\theta/2.
\end{eqnarray}
Although the coherent spin space representation involves only two variables $\theta,\phi$ rather than $2M-2$ variables $(c_j,\phi_j)$, $j=1,\ldots,M-1$, the corresponding inner product $ \langle s|{\mathbf Q}|s\rangle$ becomes a complicated function of $(\theta,\phi)$ for large $M$. This complexity carries over to the Hamiltonian of the hybrid path integral, which makes it more difficult to implement various approximation schemes. Therefore, we will work with the general spinor representation.

\subsection{Hilbert space for continuous states}

In order to incorporate spinors into the full hybrid master equation (\ref{swCK0}), we need to define corresponding operators for the continuous field $\u$. Therefore,
following Ref. \cite{Holmes20,Bressloff21b}, we introduce a Hilbert space spanned by the vectors $|\u\rangle $ together with a conjugate pair of position-momentum operators 
$\hat{u}_j,\hat{v}_j$, $j \in \calJ$, such that
\begin{equation}
\label{com2}
[\hat{u}_j,\hat{v}_k]=i \delta_{j,k},\quad [\hat{u}_j,\hat{u}_k]=0,\quad [\hat{v}_j,\hat{v}_k]=0.
\end{equation}
Their action on the given Hilbert space is taken to be
\begin{eqnarray}
{ \hat{u}_j|\u\rangle =u_j|\u\rangle ,\quad \hat{v}_j|x\rangle = -i\overset{\leftarrow}{\frac{\partial}{\partial u_j}}|\u\rangle.}
\end{eqnarray}
The arrow on the differential operator indicates that it operates to the left. Alternatively, the action of $\hat{v}_j$ can be defined in terms of state vectors,
\begin{equation}
\langle \u |\hat{v}_j|\psi\rangle =-i\frac{\partial \psi}{\partial u_j},\quad |\psi\rangle =\int d\u\, \psi(\u) |\u\rangle .
\end{equation}
The inner product and completeness relations on the Hilbert space are
\begin{equation}
\label{fpcom1}
\langle \u'|\u \rangle =\prod_{j\in \calJ}\delta(u_j-u_j'), \quad \int d\u\, |\u\rangle \langle \u|=1.
 \end{equation}
 
 It is also convenient to introduce the ``momentum'' representation
 (analogous to taking Fourier transforms),
\begin{equation}
|\v\rangle =\int d\u\,  \e^{i\v\cdot \u}|\u\rangle.
\end{equation}
It immediately follows that $|\v\rangle$ is an eigenvector of the momentum operator $\hat{v}_k$, since
\begin{eqnarray}
\hat{v}_k|\v\rangle&=\int d\u\,  \e^{i\v\cdot \u}\left (-i\overset{\leftarrow}{\frac{\partial }{\partial v_k}}\right )|\u\rangle=\int d\u\,  \e^{i\v\cdot \u} v_k|\u\rangle =v_k|\v\rangle.
\end{eqnarray}
Using the inverse Fourier transform, we also have
\begin{equation}
|\u\rangle =\int d\v \, \e^{-i\v\cdot \u}|\v\rangle,\quad \int d\v=\prod_{j\in \calJ}\int_{-\infty}^{\infty}\frac{dv_j}{2\pi},
\end{equation}
and the completeness relation
\begin{equation}
\label{fpcom2}
\int d\v |\v\rangle \langle \v|=1.
\end{equation}

\subsection{Operator version of CK equation}

Introduce the state vectors
\begin{equation}
|\psi_n(t)\rangle = \int d\u\,  \calP_n(\u,t)|\u\rangle,\quad n=1,\ldots,M.
\end{equation}
Differentiating both sides with respect to time and using equation (\ref{swCK0}) with $\Delta_{ij}^n=\Delta_{ij}$ 
gives
\begin{eqnarray}
\fl  \frac{d}{dt}|\psi_n(t)\rangle&=  \int d\u\, \bigg [-\sum_j\frac{\partial F_j(n,\u)\calP_n(\u,t)}{\partial u_j} +\sum_{m=1}^M Q_{nm} \calP_m(\u,t)\bigg ]|\u\rangle,\nonumber \\
 \fl &=\sum_{m=1}^M\left [-i\delta_{n,m}\sum_j \hat{v}_jF_{j,n}(\hat{\u})+Q_{nm}\right ] |\psi_m(t)\rangle,
\end{eqnarray}
where we have set
\begin{equation}
F_{j,n}(\u)=\sum_{k\in \calJ}\Delta_{jk}u_k +F_n(u_j).
\end{equation}
That is,
\begin{eqnarray}
\label{psiH}
\frac{d}{dt}|  {\bm \psi}(t)\rangle&=\widehat{\bf H}|  {\bm \psi}(t)\rangle,\quad |  {\bm \psi}(t)\rangle=(|  {\psi}_1(t)\rangle,\ldots,|  { \psi}_M(t)\rangle)^{\top},
\end{eqnarray} 
where
\begin{eqnarray}
\label{sh:H}
\widehat{\bf H}&=-i \sum_j \hat{v}_j \mbox{diag}(F_{j,1}(\hat{\u}),\ldots,F_{j,M}(\hat{\u}))+{\bf Q}.
\end{eqnarray}
Given the definition of the non-Hermitian Hamiltonian operator $\widehat{\bf H}$, we have 
\begin{eqnarray}
\fl  \langle s|\widehat{\bf H}|s\rangle&=H({\bf c} ,{\bm \phi},\hat{\u},\hat{\v})=-i\sum_{m=1}^M \left [\sum_{j\in \calJ}\hat{v}_jF_{j,m}(\hat{\u}) \right ]c_m +Q({\bf c},{\bm \phi}),
\label{HH}
\end{eqnarray}
with
 \begin{eqnarray}
 Q({\bf c},{\bm \phi}):= \sum_{\underset{n\neq m}{n,m=1}}^MT_{nm}\left [\e^{i(\phi_m-\phi_n)/2}-1\right ]c_m.
\end{eqnarray}

\subsection{Spatially discrete path integral}

Formally integrating equation (\ref{psiH}) yields the solution
\begin{equation}
\label{sh:sol}
|{\bm \psi}(t)\rangle=\e^{\widehat{\bf H} t}|{\bm \psi}(0)\rangle,
\end{equation}
with $\widehat{\bf H}$ given by equation (\ref{sh:H}). 
Dividing the time interval $[0,t]$ into $N$ subintervals of size $\Delta t=t/N$ then gives 
\begin{equation}
|{\bm \psi}(t)\rangle =\e^{\widehat{\bf H}\Delta t}\e^{\widehat{\bf H}\Delta t}\cdots \e^{\widehat{\bf H}\Delta t}|{\bm \psi}(0)\rangle. 
\end{equation}
Consider the product Hilbert space $|s,\u\rangle =|s\rangle \otimes \u\rangle$ and its associated completeness relation obtained by combining equations (\ref{comp2}) and (\ref{fpcom1}).  
 Introducing the integral measure
\begin{equation}
\int_{\Omega}ds = \prod_{m=1}^{M-1}\int_0^{1}\frac{dc_m}{2}\, \int_0^{4\pi}\frac{d\phi_m}{4\pi},
\end{equation}
 we insert multiple copies of the completeness relation so that
\begin{eqnarray}
   |{\bm \psi}(t)\rangle &=\int_{\Omega}ds_0 \cdots \int_{\Omega}ds_N \int d\u_0\, \cdots \int d\u_n\, 
|s_N,\u_N\rangle \nonumber \\
  &\quad \times \left [\prod_{\ell=0}^{N-1} \langle s_{\ell+1},\u_{\ell+1}|\e^{\widehat{\calH}\Delta t}|s_{\ell},\u_{\ell}\rangle\right ] \langle s_0,\u_0|{\bf \psi}(0)\rangle .
 \label{pip}
\end{eqnarray}
In the limit $N\rightarrow \infty$ and $\Delta t \rightarrow 0$ with $N\Delta t =t$ fixed, we make the approximation
\begin{eqnarray}
 \fl \langle  s_{\ell+1},\u_{\ell+1}|\e^{\widehat{\bf H}\Delta t}|   s_{\ell},\u_{\ell}\rangle &= \langle   s_{\ell+1},\u_{\ell+1}|1+\widehat{\bf H}\Delta t|   s_{\ell},\u_{\ell}\rangle+O(\Delta t^2)\\
\fl &\approx \langle s_{\ell+1} | s_{\ell} \rangle\bigg \{\delta(\u_{\ell+1}-\u_{\ell}) +\langle \u_{\ell+1}|H({\bf c}_{\ell} ,{\bm \phi}_{\ell},\u_{\ell},\hat{\v}_{\ell})\Delta t | \u_{\ell}\rangle\bigg \}.\nonumber 
\end{eqnarray}
In addition, equation (\ref{sdiff}) implies that
\begin{eqnarray}
\langle s_{\ell+1}|s_{\ell}\rangle &= 1-\frac{1}{2}i\sum_{m=1}^M (\phi_{\ell+1,m}-\phi_{\ell,m})c_{j,m}+O(\Delta \phi^2)\nonumber \\
&=1-\frac{1}{2}i\Delta t\sum_{m=1}^M\frac{d\phi_{\ell,m}}{dt}c_{\ell,m} +O(\Delta t^2).
\label{sl}
\end{eqnarray}
Each small-time propagator thus becomes
\begin{eqnarray}
\label{K1}
 &\langle s_{\ell+1},\u_{\ell+1}|\e^{\widehat{\bf H}\Delta t}| s_{\ell},\u_{\ell}\rangle \\
 &\approx \langle \u_{\ell+1}|\exp\left (\left [H({\bf c}_{\ell},{\bm \phi}_{\ell},\u_{\ell},\hat{\v}_{\ell})-\frac{i}{2}\sum_{m=1}^M \frac{d\phi_{\ell,m}}{dt}c_{\ell,m}\right ]\Delta t\right )| \u_{\ell}\rangle .\nonumber
\end{eqnarray}
Substituting the momentum completeness relation (\ref{fpcom2}) into the small-time propagator (\ref{K1}) then gives
\begin{eqnarray}
  \label{K2}
\fl &\langle s_{\ell+1},\u_{\ell+1}|\e^{\widehat{\bf H}\Delta t}| s_{\ell},\u_{j}\rangle  
\\ 
\fl &\quad  \approx \int d\v_{\ell}\langle \u_{\ell+1}| \v_{\ell}\rangle  \langle \v_{\ell}|\u_{\ell}\rangle \exp\left (\left [H({\bf c}_{\ell},{\bm \phi}_{\ell},\u_{\ell}, {\v}_{\ell})-\frac{i}{2}\sum_{m=1}^M \frac{d\phi_{\ell,m}}{dt}c_{\ell,m} \right ]\Delta t\right ).\nonumber
\end{eqnarray}
Furthermore,
\begin{eqnarray}
\label{K3}
\fl \langle \u_{\ell+1}| \v_{\ell}\rangle  \langle \v_{\ell}|\u_{\ell}\rangle & =\e^{i\v_{\ell}\cdot(\u_{\ell+1}-\u_{\ell})} =\exp\left (i\v_{\ell}\cdot \frac{d\u_{\ell}}{dt}\Delta t\right ) +O(\Delta t^2).
\end{eqnarray}

Next, substituting equations (\ref{K2}) and (\ref{K3}) into (\ref{pip}) yields 
\begin{eqnarray}
\fl & |{\bm \psi}(t)\rangle =\int_{\Omega}ds_0 \cdots \int_{\Omega}ds_N  \int d\u_0d\v_0 \cdots  \ \int d\u_Nd\v_N 
|s_N,\u_N\rangle \nonumber\\
 \fl &\quad \times \prod_{\ell=0}^{N-1} \exp\left (\left [H({\bf c}_{\ell},{\bm \phi}_{\ell},\u_{\ell}, {\v}_{\ell})-\frac{i}{2}\sum_{m=1}^M \frac{d\phi_{\ell,m}}{dt}c_{\ell,m}+i\v_{\ell}\cdot \frac{d\u_j}{dt}\right ]\Delta t\right )\langle s_0,\u_0|{\bm \psi}(0)\rangle . \nonumber
 \label{pip2}
\end{eqnarray}
The final step is to take the limit $N\rightarrow \infty,\Delta t\rightarrow 0$ with $N\Delta t=t$ fixed, $\u_{\ell}=\u(\ell\Delta t)$ etc. We will also assume that $<\u_0|\psi_n(0)\rangle =\rho_n \delta(\u-\u_0)$,
and set
\[P_n(\u,t|\u_0,0)=\langle \u,n|{\bm \psi}(t)\rangle.\]
After Wick ordering, $\v\rightarrow -i\v$ and integrating by parts the term involving $d\phi/dt$, we obtain the following functional path integral:
\begin{eqnarray}  & P_n(\u,t|\u_0,0)={\mathcal N}_n \int_{\u(0)=\u_0}^{\u(t)=\u} {\mathcal D}[{\bf c}]  {\mathcal D}[{\bm \phi}]{\mathcal D}[\u]{\mathcal D}[\v]\nonumber \\
\fl &\qquad \qquad \times  \exp\left (-\int_{0}^{t}\left [\v\cdot \frac{d\u}{d\tau}-\frac{i}{2}\sum_{m=1}^M\phi_m\frac{dc_m}{d\tau} -{\mathcal H}\right ]d\tau\right ), 
\label{dpi}
\end{eqnarray}
where ${\mathcal H}$ is the effective Hamiltonian 
\begin{eqnarray}
\fl  {\mathcal H}&= \sum_{m=1}^M \left [\sum_{i,j\in \calJ}\bigg (v_j \Delta_{jk}u_k +F_m(u_j) \bigg )\right ]c_m +\sum_{\underset{n\neq m}{n,m=1}}^MT_{nm}\left [\e^{i(\phi_m-\phi_n)/2}-1\right ]c_m,
\label{Heff0}
\end{eqnarray}
with ``position coordinates'' $(\u,{\bf c})$ and ``conjugate momenta'' $(\v,-i{\bm \phi})$. Here ${\mathcal N}_n$ is a normalization constant. (Note that there are  $M-1$ independent auxiliary coordinates, $c_1,\ldots,c_{M-1}$ due to the normalization condition $\sum_{m=1}^Mc_m=1$.)

\setcounter{equation}{0}
\section{Continuum path integral and the weak-noise limit}

Having obtained the path integral of the spatially discretized hybrid system evolving according to equation (\ref{swCK0}), we can now take the continuum limit to determine the corresponding functional path integral for the original stochastic hybrid diffusion equation (\ref{swdiff}). For the sake of generality, we will write down the result for the $d$-dimensional version of the stochastic hybrid RD equation, which takes the form
\begin{equation}
\frac{\partial U}{\partial t}=D\nabla^2 U+F_n(U).
\label{hdRD}
\end{equation}
That is, the derivation of the path integral (\ref{dpi}) carries over straightforwardly to the spatially discretized version of (\ref{hdRD}): the only change is that $\Delta_{ij}$ becomes the discrete Laplacian on a $d$-dimensional square lattice. The resulting  continuum path integral in $d$ spatial dimensions is
\begin{eqnarray}
  P_n[u]={\mathcal N}_n \int_{u(x,t)=u(x)} {\mathcal D}[{\bf c}]  {\mathcal D}[{\bm \phi}]{\mathcal D}[u]{\mathcal D}[v] \e^{-S[u,v,{\bf c},{\bm \phi}]},  
\label{pieff}
\end{eqnarray}
with the action functional
\begin{eqnarray}
\label{actRD0} 
\fl S&=\int_0^t \bigg \{ 
\int_{\R^d} v\left (\frac{\partial u}{\partial \tau}-D\nabla^2u\right ) d\x   -\sum_{m=1}^M c_m \int_{\R^d}v F_m(u) d\x  -\frac{i}{2} \sum_{m=1}^M\phi_m\frac{dc_m}{d\tau} \nonumber \\
 \fl &\qquad  -\sum_{\underset{n\neq m}{n,m=1}}^MT_{nm}\left [\e^{i(\phi_m-\phi_n)/2}-1\right ]c_m \bigg\}d\tau  .
\end{eqnarray}

\subsection{Semi-classical limit}

One of the useful features of path-integral representations is that they provide a systematic framework for developing various approximation schemes, including diagrammatic perturbation theory, renormalization group theory and effective actions, and weak-noise approximations \cite{Zinn02,Kleinert09,Chow15}. We will focus on the latter here. As a first step, consider the scalings ${\bf T} \rightarrow {\bf T}/\epsilon$ and $v\rightarrow v/\epsilon$, and rewrite the path integral (\ref{pieff}) as
\begin{eqnarray}
  P_n[u]={\mathcal N}_n \int_{u(x,t)=u(x)} {\mathcal D}[{\bf c}]  {\mathcal D}[{\bm \phi}]{\mathcal D}[u]{\mathcal D}[v] \e^{-S[u,v,{\bf c},{\bm \phi}]/\epsilon}, \nonumber \\
\label{piegw}
\end{eqnarray}
with the action
\begin{eqnarray}
\fl S&=\int_0^t \bigg \{ -\frac{i\epsilon}{2} \sum_{m=1}^M\phi_m\dot{c}_m-\sum_{\underset{n\neq m}{n,m=1}}^MT_{nm} \left [\e^{i (\phi_m-\phi_n)/2}-1\right ]c_m +\int_{\R^d}  l(u,v,{\bf c})d\x\bigg\}d\tau .\nonumber \\
\fl 
\label{actRD2}
\end{eqnarray}
Here
\begin{eqnarray}
l(u,v,{\bf c})&= v\left (\frac{\partial u}{\partial \tau}-D\nabla^2u \right )  -v\sum_{m=1}^Mc_m F_m(u).
\end{eqnarray}
In order to derive a least-action principle for the concentration field $u(\x)$, we successively eliminate the auxiliary variables ${\bm \phi}$ and ${\bf c}$ in the limit $\epsilon \rightarrow 0$. First, to leading order we can drop the $O(\epsilon)$ term in the action (\ref{actRD2}). Hence, the only dependence on the momentum variables $\phi_m$ is via the term
\begin{equation}
{\mathcal T}\equiv \sum_{\underset{n\neq m}{n,m=1}}^MT_{nm} \left [1-\e^{i (\phi_m-\phi_n)/2}\right ]c_m .
\end{equation}
It is convenient to replace the transition matrix ${\bf T}$ on the right-hand side by the matrix generator ${\bf Q}$, which is allowed since the contribution from the diagonal part of ${\bf Q}$ vanishes. Defining $z_n=\e^{-i\phi_n/2}$, we have
\begin{equation}
{\mathcal T}= \sum_{n,m=1}^MQ_{nm} \left [1-\frac{z_n}{z_m}\right ]c_m.
\label{calT}
\end{equation}
We will assume that we can Wick rotate the momentum variables $\phi_n$ such that $z_n$ becomes positive and real. Minimizing the action with respect to the ${\bm \phi}$ then reduces to the problem of minimizing ${\mathcal T}$ with respect to $\z$ for fixed ${\bf c}$. We will proceed by adapting arguments developed in \cite{Bressloff17}.

First, consider the ansatz that the solution of the variational problem for $\z$ is given by the eigenvector of the following linear equation:
\begin{equation}
\label{zdual}
q_mz_m+\sum_{n=1}^MQ_{nm} z_n=\lambda z_m
\end{equation}
for some bounded vector  ${\bf q}=(q_1,\ldots,q_M)$.
Since $\sum_{m=1}^MQ_{nm}=0$, it follows that we are free to shift the vector $\q$ by a constant. In other words, we can set $q_M=0$, say, and consider the $M-1$ independent variables $q_1,\ldots q_{M-1}$. The Perron-Frobenius theorem ensures that there exists a unique positive solution $z_n=z_n(\q)$, $\q=(q_1,\ldots,q_{M-1},0)$, with $\lambda(\q)$ the Perron or principal eigenvalue and the normalization condition $\sum_{m=1}z_m=1$. Combining equation (\ref{zdual}) with (\ref{calT}) then gives
\begin{equation}
{\mathcal T} = \sum_{m=1}^M(q_m-\lambda)c_m=\sum_{m=1}^{M-1}q_mc_m-\lambda,
\end{equation}
since $\sum_{m=1}^Mc_m=1$ and $q_M=0$.
It also follows that the $O(1)$ form of the action (\ref{actRD2}) becomes
\begin{eqnarray}
S&=\int_0^t \bigg \{ \sum_{m=1}^{M-1}q_m c_m-\lambda(\q)+\int_{\R^d}  l(u,v,{\bf c})d\x\bigg\}d\tau ,\label{actRD3}
\end{eqnarray}
and the variational equation with respect to ${\bf z}$ reduces to
\begin{equation}
0=\frac{\delta S}{\delta z_n}=\sum_{m=1}^{M-1} \frac{\partial q_m}{\partial z _n}\left [c_m - \frac{\partial \lambda}{\partial q_m} \right ].\end{equation}
Hence, the Perron eigenvector $\z(\q)$ solves the given variational problem provided that 
\begin{equation}
c_m=c_m(\q)\equiv \frac{\partial \lambda}{\partial q_m}  \quad n=1,\ldots,M-1 .
\label{clam}
\end{equation}

It remains to show that $c_m(\q)$, $m=1,\ldots,M$, exists and that the inverse functions $q_m=q_m({\bf c})$ also exist. This can be achieved by considering the unique positive eigenvector ${\bf R}(\q)$ (up to scalar multiplication) of the adjoint linear equation
\begin{equation}
\label{Rdual}
q_mR_m+\sum_{n=1}^MQ_{mn} R_n=\lambda R_m.
\end{equation}
Differentiating equation (\ref{Rdual}) with respect to $q_n$ gives
\begin{equation}
R_n\delta_{n,m}+q_m\frac{\partial R_m}{\partial q_n}+\sum_{l=1}^MQ_{ml} \frac{\partial R_l}{\partial q_n}=\frac{\partial \lambda}{\partial q_n} R_m+\frac{\partial R_m}{\partial q_n} \lambda.
\end{equation}
Multiplying both sides of this equation by $z_m $ and summing over $m$ yields
\begin{equation}
\fl z_nR_n+\sum_{m=1}^M z_m q_m\frac{\partial R_m}{\partial q_n}+\sum_{l=1}^Mz_mQ_{ml} \frac{\partial R_l}{\partial q_n}=\frac{\partial \lambda}{\partial q_n} \sum_{m=1}^Mz_mR_m+\lambda \sum_{m=1}^M z_m\frac{\partial R_m}{\partial q_n} .
\end{equation}
Now imposing equation (\ref{zdual}) shows that most terms cancel, resulting in the simple relation 
\begin{equation}
c_n(\q)=\frac{\partial \lambda}{\partial q_n}=R_n(\q)z_n(\q), 
\label{clam2}
\end{equation}
after imposing the normalization $ \sum_{m=1}^M R_n(\q)z_n(\q)=1$.
Since ${\bf R}$ and $\z$ are strictly positive, $\lambda(\q)$ is a monotonically increasing function of the $q_m$. Moreover, equations (\ref{zdual}) and (\ref{Rdual}) imply that in the limit $q_l\rightarrow \infty$ with all other components finite, $R_l,z_l\rightarrow 1$. Conversely, if $q_l\rightarrow -\infty$ then $R_l,z_l\rightarrow 0$. Hence, the range of the monotonically increasing function $\partial \lambda /\partial q_l$ is the unit interval, which means that for a given vector ${\bf c}$, there exists a vector $\q$ such that 
$c_n=R_n(\q)z_n(\q)$. It turns out that this solution is unique, since it can be shown that the Hessian matrix with elements $H_{nm}=\partial^2 \lambda/\partial q_nq_m$ is invertible \cite{Bressloff17}. Uniqueness then follows from the inverse function theorem.

Having eliminated the auxiliary momentum variables, we obtain the reduced path integral
\begin{eqnarray}
 P_n[u]\approx {\mathcal N}_n \int_{u(x,t)=u(x)} {\mathcal D}[{\bf c}]  {\mathcal D}[u]{\mathcal D}[v] \e^{-\widehat{S}/\epsilon},
\label{pired}
\end{eqnarray}
with
\begin{eqnarray}
 \widehat{S}= \int_{0}^{t}\bigg [  \sum_{m=1}^{M-1} c_mq_m -\lambda +\int_{\R^d} l(u,v,{\bf c})d\x  \bigg ]d\tau . 
\label{act33}
\end{eqnarray}
We can now eliminate the auxiliary coordinates ${\bf c}$ by functionally minimizing the action $\widehat{S}$ with respect to ${\bf c}$, noting that $q_m$ and $\lambda$ are functions of ${\bf c}$:
\begin{eqnarray}
\label{dude}
 \fl  0&=\frac{\delta \widehat{S}}{\delta c_n}= \sum_{m=1}^{M-1}\frac{\partial q_m}{\partial c_n}c_m+q_n\delta_{n,m}-\sum_{m=1}^{M-1}\frac{\partial \lambda}{\partial q_m }\frac{\partial q_m}{\partial c_n} -\delta_{n,m}\int_{\R^d}v F_n(u)d\x .
\end{eqnarray}
It then follows from equation (\ref{clam2}) that
\begin{equation}
q_n=\int_{\R^d} v F_n(u) d\x .
\end{equation}
Finally, substituting for $q_m$ in equations (\ref{zdual}) and (\ref{Rdual}) yields the following path-integral for small $\epsilon$:
\begin{eqnarray}
 & P[u]\sim \int \limits_{u(\x,t)=u(\x)}  {\mathcal D}[v]{\mathcal D}[u]  \e^{- {S[u,v]}/{\epsilon}} ,
 \label{path}
\end{eqnarray}
where $S$ is the effective action functional
\begin{eqnarray}
\label{S3}
 \fl S[u,v]&=\int_{0}^{t} \bigg \{\int_{\R^d}  v(\x,\tau)\left [ \frac{\partial}{\partial \tau}u(\x,\tau)-D\nabla^2 u(\x,\tau) \right ]d\x -\Lambda[u,v]\bigg\}d\tau,
\end{eqnarray}
and $\Lambda$ is the Perron or principal eigenvalue of the functional eigenvalue equation
\begin{eqnarray}
 \fl &\sum_{m=1}^M\left \{\delta_{m,n}  \int_{\R^d} v(\x)F_n(u(\x))d\x+Q_{nm}\right \}R_{m}[u,v] = {\Lambda}[u,v]R_{n}[u,v],
\label{FTeigR}
\end{eqnarray}
and its adjoint
\begin{eqnarray}
 \fl &\sum_{m=1}^M\left \{\delta_{m,n}  \int_{\R^d} v(\x)F_n(u(\x))d\x+Q_{nm}\right \}z_{n}[u,v] = {\Lambda}[u,v]z_{n}[u,v],
\label{FTeigz}
\end{eqnarray}
These are supplemented by the normalization conditions
\begin{equation}
\label{norm}
\sum_{m=1}^M R_{n}[u,v]z_{n}[u,v]=1,\quad \sum_{m=1}^Mz_{n}[u,v]=1.
\end{equation}

We can interpret $S[u,v]$ as the action functional of an effective Hamiltonian system with conjugate fields $u,v$ and Hamiltonian 
\begin{equation}
\label{ham}
H[u,v]=D\int_{\R^d} v(\x) \nabla^2 u(\x)d\x+\Lambda[u,v].
\end{equation}
Least action paths in the limit $\epsilon \rightarrow 0$ then correspond to solutions of Hamilton's equations
\begin{eqnarray}
\frac{\partial u(\x,t)}{\partial t}&=D \nabla^2 u(\x,t) +\frac{\delta \Lambda[u,v]}{\delta v(\x,t)},\\ \frac{\partial v(\x,t)}{\partial t}&=-D \nabla^2 v(\x,t) -\frac{\delta \Lambda[u,v]}{\delta u(\x,t)}.
\end{eqnarray}
Using Hamilton-Jacobi theory, it follows that evaluating the action along a ``zero energy'' least action path gives $S_{\rm opt}=\Phi[u]$,
where $\Phi$ is a solution of the Hamilton-Jacobi equation
\begin{equation}
H[u,\delta_u\Phi]=0,\quad \delta_u\Phi=\frac{\delta \Phi}{\delta u(\x)}.
\label{HJ}
\end{equation}
One solution is $v=\delta_u\Phi=0$ for which $\Lambda[u,0]=0$, $\delta\Lambda[u,0]/\delta u(\x)=0$ and $R_n[u,0]=\rho_n$ with $\rho$ the stationary distribution of the Markov chain. Functionally differentiating the eigenvalue equation (\ref{FTeigR}) with respect to $v(\x)$ shows that
\begin{eqnarray}
\fl  &\sum_{m\geq 0}\left \{ \int_{\R^d} d\x \, v(\x)F_n(u(\x))\delta_{m,n} +Q_{nm}\right \}\frac{\delta R_{m}}{\delta v(\x)} +F_n(u(\x)) R_n = {\Lambda} \frac{\delta R_{n}}{\delta v(\x)}+\frac{\delta \Lambda}{\delta v(\x)}R_n .\nonumber \\
\fl
 \label{bog}
\end{eqnarray}
Summing both sides with respect to $n$ and setting $v=0$ with $\sum_nQ_{nm}=0$ then gives
\begin{equation}
\sum_n \rho_nF_n(u(\x)) =\left. \frac{\delta\Lambda[u,v]}{\delta v(\x)}\right |_{v=0}.
\end{equation}
Hence, along the least action path, $u(\x,t)$ evolves according to the deterministic mean field equation
\begin{eqnarray}
  \frac{\partial u}{\partial t}&=D\nabla^2 u +\overline{F}(u),\quad  \overline{F}(u(\x))=\sum_n \rho_nF_n(u(\x)) .
  \label{Fbar}
\end{eqnarray}

Equation (\ref{HJ}) is the functional analog of the Hamilton-Jacobi equation previously derived for finite-dimensional stochastic hybrid systems \cite{Bressloff13a,Bressloff14,Bressloff15}. For example, consider the piecewise deterministic equation
\begin{equation}
\frac{du}{dt}=F_n(u),\quad u \in \R
\end{equation}
for $N(t)=n$, with $N(t)$ evolving according to a continuous-time Markov chain with generator ${\bf Q}$. The corresponding hybrid path integral takes the form
\begin{eqnarray}
 & P \sim \int \limits_{u(t)=u}  {\mathcal D}[v]{\mathcal D}[u]  \exp\left (- \epsilon^{-1}\int_{0}^{t} \bigg (v\dot{u}-\Lambda(u,v)\bigg )d\tau \right ) ,
 \label{pathfin}
\end{eqnarray}
where $\Lambda$ is the Perron eigenvalue of the linear equation
\begin{eqnarray}
\sum_{m=1}^M\left (\delta_{m,n}   v F_n(u ) +Q_{nm}\right )R_{m}(u,v)= {\Lambda}(u,v)R_{n}(u,v),
\label{eigR}
\end{eqnarray}
with $\sum_{m=1}^M R_{n} z_{n} =1$ and $z_n$ the adjoint eigenvector. Suppose that the corresponding mean-field equation
\begin{equation}
\frac{du}{dt}=\overline{F}(u)\equiv \sum_{n=1}^M\rho_n F_n(u)
\end{equation}
has a pair of stable fixed points separated by an unstable fixed point. In the weak noise limit, it can be shown that the optimal (most likely) paths of escape from one of the metastable states are given by the non-classical zero energy solutions of the Hamilton-Jacobi equation \cite{Bressloff13a,Bressloff14,Bressloff15}
\begin{equation}
\Lambda(u,\partial_u\Phi)=0.
\end{equation}
In addition, the resulting Arrhenius factor in the expression for the mean escape rate involves an effective potential that is consistent with WKB methods and the rate function of large deviation theory \cite{Kifer09,fagg09,Faggionato10,Bressloff17}. Combining these observations with previous studies of stochastic PDEs \cite{Maier01,Stein04,Stein05} suggests that non-classical solutions of the Hamilton-Jacobi functional equation (\ref{HJ}) could be used to study metastability in a stochastic hybrid RD model.

\subsection{Gaussian approximation}

In cases where rare events do not play a significant role, we can use the semi-classical path integral (\ref{path}) to derive a Gaussian approximation of the stochastic hybrid RD equation (\ref{hdRD}). We proceed along similar lines to finite-dimensional hybrid systems \cite{Bressloff21a} by carrying out a ``small momentum expansion'' with respect to $v$. Introduce the rescaling $v\rightarrow \epsilon v$
and consider the expansion of the functional eigenvalue $\Lambda$ 
and the associated eigenvector ${\bf R}$:
\numparts
\begin{eqnarray}
\fl & R_n[u,\epsilon v]=R_{n,0}[u]+\epsilon\int_{\R^d} v(\x) R_{n,1}(u(\x))d\x+\epsilon^2 \int_{\R^d} v^2(\x) R_{n,2}(u(\x))d\x+\ldots,\\
\fl &\Lambda[u,\epsilon v]=\Lambda_0[u]+\epsilon \int_{\R^d} v(\x) \Lambda_{1}(u(\x))d\x+\epsilon^2   \int_{\R^d} v^2(\x) \Lambda_{2}(u(\x))d\x+\ldots,
\end{eqnarray}
\endnumparts
 Substituting into the eigenvalue equation (\ref{FTeigR}),
 \begin{eqnarray}
\fl & \sum_{m=1  }^N\left [Q_{nm} +  \epsilon \delta_{n,m}v\circ F_n(u)\right ]\bigg (R_{m,0}[u]+\epsilon v\circ R_{n,1}(u)+\epsilon^2 v^2\circ  R_{n,2}(u)+\ldots\bigg )\nonumber\\
\fl&\quad =\left [ \Lambda_0[u]+\epsilon v\circ \Lambda_1(u)+\epsilon^2 v^2 \circ \Lambda_2(u)+\ldots\right ]\nonumber\\
\fl &\hspace{4cm} \times \bigg (R_{m,0}[u]+\epsilon v\circ R_{n,1}(u)+\epsilon^2 v^2\circ  R_{n,2}(u)+\ldots\bigg ).
\end{eqnarray}
We have introduced the compact notation 
\[v^k\circ f(u)=\int_{\R^d} v^k(\x)f(u(\x))d\x.\]
Collecting terms in equal powers of $\epsilon$ yields a hierarchy of equations. The first three are
\numparts
\begin{eqnarray}
\fl &\sum_{m }\bigg (Q_{nm}-\Lambda_0[u]\delta_{n,m}\bigg )R_{m,0}[u]=0,\\
\fl &\sum_{m  }\bigg (Q_{nm} -\Lambda_0[u]\delta_{n,m}\bigg )v\circ R_{m,1}(u)=-v\circ F_n(u) R_{n,0}[u] +v\circ \Lambda_1[u]R_{n,0}[u],\\
\fl &\sum_{m  }\bigg (Q_{nm}-\Lambda_0[u]\delta_{n,m}\bigg )v\circ R_{m,2}(u) =-\big [v\circ F_n(u) \big ]\big [v\circ R_{n,1}(u)\big ]\\
\fl &\hspace{6cm}+\big [v\circ \Lambda_1(u)\big ]\big [v\circ R_{n,1}(u)\big ]+ v\circ \Lambda_2(u) R_0[u].\nonumber
\end{eqnarray}
\endnumparts
The first equation has the solution $\Lambda_0[u]=0$ and $R_{m,0}=\rho_m$ independently of $u$.
Applying the Fredholm alternative theorem to the second and third equations by summing over $n$ gives the self-consistency conditions
\numparts
\begin{eqnarray}
\fl 0&=\sum_n\bigg \{-v\circ F_n(u) R_{n,0}[u] +v\circ \Lambda_1(u)R_{n,0}[u]\bigg \},\\
\fl 0&=\sum_n\bigg \{-\big [v\circ F_n(u) \big ]\big [v\circ R_{n,1}(u)\big]+\big [v\circ \Lambda_1(u)\big ]\big [v\circ R_{n,1}(u)\big ]+ v\circ \Lambda_2(u) R_0[u].\bigg \}.\nonumber \\
\fl
\end{eqnarray}
\endnumparts
The normalization conditions (\ref{norm}) imply that $\sum_{m=1}^M R_{n,k}=\delta_{k,0}$, which leads to the results
\numparts
\begin{eqnarray}
v\circ \Lambda_1(u)&=\sum_{n=1}^M\rho_n v\circ F_n(u),\\
v\circ \Lambda_2(u)&=\sum_{n=1}^M \big [v\circ F_n(u) \big ]\big [v\circ R_{n,1}(u)\big].
\end{eqnarray}
\endnumparts

Ignoring higher-order terms, we thus have the following Gaussian approximation of the principal eigenvalue:
\begin{equation}
\label{lamexp}
\Lambda[u,\epsilon v]\approx \epsilon v \circ \overline{F}(u)+\epsilon^2 \sum_{n=1}^M \big [v\circ F_n(u) \big ]\big [v\circ Z_n(u)\big],
\end{equation}
with $\overline{F}(u)$ defined in equation (\ref{Fbar}) and $Z_n(u)\equiv R_{n,1}(u)$ satisfying the linear equation
\begin{equation}
\label{Z}
\sum_{m =1}^M Q_{nm}  Z_m(u)=  [\overline{F}(u)-F_n(u)]\rho_n,\quad \sum_{n=1}^MZ_n(u)=0.
\end{equation}
Note that a unique solution for $Z_m(u)$ exists even though the matrix ${\bf Q}$ is singular, which is a consequence of the Fredholm alternative theorem. Substituting the Gaussian approximation into the action functional (\ref{S3}) gives
\begin{eqnarray}
\label{Sgauss}
 \fl S[u,v]&=\int_{0}^{t} \bigg \{\int_{\R^d}  v(\x,\tau)\left [ \frac{\partial}{\partial \tau}u(\x,\tau)-D\nabla^2 u(\x,\tau) -\overline{F}(u(\x,\tau))\right ]d\x \nonumber \\
\fl  &\qquad +\int_{\R^d} \int_{\R^d}  v(\x,\tau) v(\x',\tau)C(u(\x,\tau),u(\x',\tau))\bigg\}d\tau,
\end{eqnarray}
with
\begin{equation}
C(u(\x),u(\x'))=\sum_{n=1}^MF_n(u(\x))Z_n(u(\x')).
\end{equation}

Having reduced the action to a quadratic in the momentum variables, it follows that the corresponding path integral represents an equivalent SDE  
with Gaussian spatiotemporal noise that is white with respect to time and colored with respect to space. Since the quadratic term in the action depends on the concentration, the corresponding noise is multiplicative, which means that there is an ambiguity in the interpretation of the noise in the sense of Ito vs. Stratonovich. One way to resolve this issue would be to calculate any contributions to the functional measure $ {\mathcal D}[v]{\mathcal D}[u] $ in equation (\ref{path}) arising from the elimination of the auxiliary variables. However, since changing the interpretation simply generates an $O(\epsilon)$ correction to the deterministic part of the dynamics, we will consider the Ito version here. Introducing the stochastic concentration field $U(\x,t)$, the Ito SDE takes the form
\begin{equation}
\label{Usde}
\frac{\partial U(\x,t)}{\partial t} =D\nabla^2 U(\x,t)+\overline{F}(U(\x,t))+\sqrt{2\epsilon}\xi(\x,t),
\end{equation}
with $\langle \xi(\x,t)\rangle =0$ and
\begin{equation}
\langle \xi(\x,t)\xi(\x',t')\rangle = \delta(t-t')C(U(\x,t),U(\x',t)).
\end{equation}

\section{Two-dimensional diffusion on a switching substrate} 

\begin{figure}[b!]
 \raggedleft
\includegraphics[width=10cm]{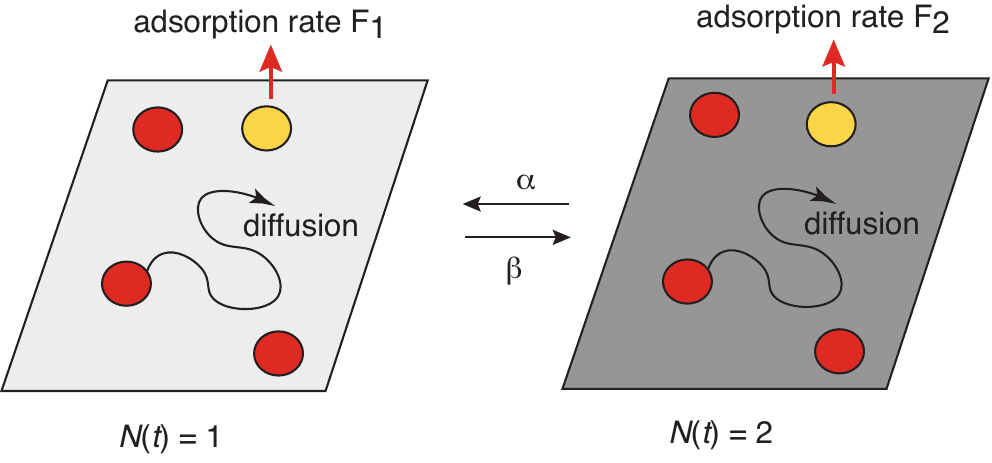}
\caption{Schematic diagram of diffusion over a stochastically switching two-dimensional substrate. The substrate switches between two discrete states according to a Markov chain with transition rates $\alpha,\beta$. Particle reactions are catalyzed in the active state. Reactions could involve single particle adsorption (as shown in the figure) or a binary reaction such as pair annihilation, for example.} \label{fig1}
\end{figure}

As an illustration of the above theory, consider diffusion over a two-dimensional substrate that switches between an active state $(n=1)$ and an inactive state ($n=2)$; chemical reactions are enhanced in the former case, see Fig. \ref{fig1}. Examples of chemical reactions could include adsorption by the substrate \cite{Bressloff17b}, or a binary reaction such as pair annihilation. Suppose that the substrate switches between the two states according to the two-state Markov chain 
\[
 1\Markov{\alpha}{\beta}2.\]
 The corresponding matrix generator is 
 \begin{equation}
 \label{Q}
 {\bf Q}=\left (\begin{array}{cc} -\beta &\alpha \\ \beta  &-\alpha\end{array}\right ),
\end{equation}
 and the steady-state distribution of the master equation (\ref{master}) is
 \begin{equation}
\rho_1=\frac{\alpha}{\alpha+\beta},\quad \rho_2=\frac{\beta}{\alpha+\beta}.
\end{equation}
Let $N(t)\in\{1,2\}$ denote the current state of the substrate. The stochastic hybrid RD equation is then
 \begin{equation}
\frac{\partial U}{\partial t}=D\nabla^2 U+F_n(U),\quad \x \in \R^2,
\label{hdRD2}
\end{equation}
for $N(t)=n$. 

\subsection{Path integral in the semi-classical limit}

In order to determine the effective action (\ref{S3}) for the path integral (\ref{path}), it is necessary to solve the functional eigenvalue equation (\ref{Rdual}), which takes the particular form
\begin{equation}
  \left (\begin{array}{cc} -\beta+ v\circ F_1(u) & \alpha\\ \beta & -\alpha+ v\circ F_2(u)\end{array}\right )\left (\begin{array}{c} R_1 \\ R_2\end{array}\right )=\Lambda \left (\begin{array}{c} R_1 \\ R_2\end{array}\right ).
\end{equation}
The corresponding characteristic equation is
\begin{eqnarray}
\fl \Lambda^2+\Lambda[\alpha+\beta -  v\circ [F_1(u)+F_2(u)]+(\alpha-v\circ F_2(u)) (\beta-  v\circ F_1(u))-\alpha \beta=0.
\end{eqnarray}
It follows that the principal eigenvalue is given by
\begin{eqnarray}
\Lambda[u,v]=\frac{1}{2}\left [  \Sigma[u,v]+\sqrt{\Sigma[u,v]^2- 4h[u,v]}    \right ],
\end{eqnarray}
where
\numparts
\begin{eqnarray}
\Sigma[u,v]&=  v\circ  [F_1(u)+F_2(u)]-(\alpha+\beta),\\
h[u,v])&=- v\circ[ \beta F_2(u)+\alpha F_1(u)]+ [ v\circ  F_1(u)][ v\circ  F_2(u)].
\end{eqnarray}
\endnumparts
It can be checked that the discriminant is positive so that $\Lambda[u,v]$ is real. Substituting for $\Lambda$ into equations (\ref{ham}) and (\ref{HJ}) leads to the following Hamilton-Jacobi equation for
the quasipotential $\Phi$:
\begin{equation}
\label{HJab}
\fl D\int_{\R^d} \frac{\delta \Phi}{\delta u(x)} \nabla^2 u(x)dx+\frac{1}{2}\left [  \Sigma[u,\delta_u\Phi]+\sqrt{\Sigma[u,\delta_u\Phi]^2- 4h[u,\delta_u\Phi]}\right ]=0.
\end{equation}
The trivial solution of this equation is $\delta_u\Phi=0$, whose least-action path represents a solution to the deterministic mean field equation
\begin{equation}
\label{mf}
\frac{\partial u}{\partial t}=D\nabla^2 u+\overline{F}(u),\quad \overline{F}(u)=  \rho_1 F_1(u)+\rho_2 F_2(u).
\end{equation}
There may also exist nontrivial solutions of the Hamilton-Jacobi equation that are spatially uniform. Equation (\ref{HJab}) now reduces to the form
\begin{equation}
\frac{1}{2}\left [  \Sigma(u,\Phi')+\sqrt{\Sigma(u,\Phi')^2- 4h(u,\Phi')}\right ]=0,\quad \Phi'=\frac{d\Phi}{du},
\end{equation}
which requires
\begin{equation}
h(u,\Phi')=0,\quad \Sigma(u,\Phi') <0.
\end{equation}
Substituting for $h$ yields the quadratic equation
\begin{equation}
 \Phi'\bigg [ \beta F_2(u)+\alpha F_1(u)- \Phi' F_1(u)   F_2(u)\bigg ]=0,
 \end{equation}
 with the solutions $\Phi'=0$ and
 \begin{equation}
 \label{qua}
 \Phi'(u)=\frac{\beta}{ F_1(u)}+\frac{\alpha}{ F_2(u)}.
 \end{equation}
 The latter will generate a quasipotential along a non-deterministic path, provided that the functions $F_n(U)$, $n=1,2$, are non-vanishing. (One application of this non-spatial, two-state model is to a bistable genetic switch; equation (\ref{qua}) then determines the action along the most likely path of escape from one of the two metastable states \cite{Bressloff17a}.) 
A challenging mathematical problem is finding nontrivial spatially varying solutions of (\ref{HJab}) and interpreting them in terms of large deviation theory.

\subsection{Gaussian approximation}

In order to determine the Gaussian correction to the mean-field equation (\ref{mf}) we have to solve equation (\ref{Z}), which becomes
\begin{equation}
  \left (\begin{array}{cc} -\beta& \alpha\\ \beta & -\alpha \end{array}\right )\left (\begin{array}{c} Z_1 \\ Z_2\end{array}\right )=\left (\begin{array}{c}   -\rho_1\rho_2 [F_1(u)-F_2(u)] \\  \rho_1\rho_2 [F_1(u)-F_2(u)]\end{array}\right ) .
\end{equation}
Since $Z_1+Z_2=0$, we obtain the unique solution
\begin{equation}
Z_1=-Z_2= \frac{ \rho_1\rho_2}{\alpha+\beta} [F_1(u)-F_2(u)].
\end{equation}
The resulting Ito SDE (\ref{Usde}) thus takes the form
\begin{equation}
\label{Usde2}
\frac{\partial U(x,t)}{\partial t} =D\nabla^2 u+\overline{F}(U(x,t))+\sqrt{2\epsilon}\xi(x,t),
\end{equation}
with $\langle \xi(x,t)\rangle =0$,
\begin{equation}
\langle \xi(x,t)\xi(x',t')\rangle = \delta(t-t')C(U(x,t),U(x',t)) ,
\end{equation}
and
\begin{equation}
\label{B1}
\fl C(U(x,t),U(x',t))=  \frac{\rho_1\rho_2}{\alpha+\beta}[F_1(U(x,t))-F_2(U(x,t))] [F_1(U(x',t))-F_2(U(x',t))] .
\end{equation}

Taking expectations of both sides of the SDE (\ref{Usde2}) gives
\begin{eqnarray}
  \frac{\partial \langle U\rangle }{\partial t}&=D\nabla^2 \langle U \rangle+\langle \overline{F}(U)\rangle .
\end{eqnarray}
Let $u$ be the solution to the deterministic RD equation (\ref{mf}) and introduce the linear noise approximation
\begin{equation}
U(x,t)=u(x,t)+\sqrt{\epsilon}\widetilde{U}(x,t).
\end{equation}
We then have
\begin{eqnarray*}
\fl  \frac{\partial \langle U\rangle }{\partial t}&=D\nabla^2  \langle U\rangle  +\langle \overline{F}(u+\sqrt{\epsilon}\widetilde{U})\rangle 
 =D\nabla^2  \langle U\rangle+\langle \overline{F}(\langle U\rangle +\sqrt{\epsilon}(\widetilde{U}-\langle \widetilde{U}\rangle)\rangle\\
 \fl &\approx D\nabla^2  \langle U\rangle+ \overline{F}(\langle U\rangle )+\sqrt{\epsilon}\overline{F}'(\langle U\rangle )\langle (\widetilde{U}-\langle \widetilde{U}\rangle)\rangle +\frac{\epsilon}{2} \overline{F}''(\langle U\rangle )\langle (\widetilde{U}-\langle \widetilde{U}\rangle)^2\rangle.
\end{eqnarray*}
The leading order correction to equation (\ref{mf}) is thus as follows:
\begin{equation}
\label{pyg2}
\frac{\partial\langle U\rangle}{\partial t}=D\nabla^2 \langle U\rangle+\overline{F}(\langle U\rangle)+\frac{\epsilon}{2}\Delta(x,x,t)\overline{F}''(\langle U\rangle ),
\end{equation}
where $\Delta(x,y,t)$ is the equal-time two-point correlation function
\begin{equation}
\Delta(x,y,t)=\langle \delta \widetilde{U}(x,t)\delta \widetilde{U}(y,t)\rangle,
\end{equation}
with $\delta \widetilde{U}=\widetilde{U}-\langle \widetilde{U} \rangle$. 

Since the first-order moment equation couples to the two-point correlation function at $O(\epsilon)$ when $F(U) $ is a nonlinear function of $U$, we need to obtain an equation for the latter. Again this can be obtained using the linear noise approximation. More specifically, linearizing equation (\ref{Usde2}) about $u$ gives
\begin{eqnarray}
\label{swgauss2}
  \frac{\partial \widetilde{U}}{\partial t}&=D\nabla^2 \widetilde{U}+\overline{F}'(u)\widetilde{U}+ \widetilde{\xi}(x,t),
\end{eqnarray}
with $\langle \widetilde{\xi}(x,t)\rangle =0$ and
\begin{equation}
\label{xi2}
\langle \widetilde{\xi}(x,t)\widetilde{\xi}(x',t)\rangle =2  C(u(x,t),u(x',t')) \delta(t-t').
\end{equation}
We can calculate moments of the stochastic density $\widetilde{U}(x,t)$ by expressing the solution in terms of the causal Green's function or propagator $G$, which is defined according to
\begin{eqnarray}
\label{GG}
&\left (\frac{\partial}{\partial t}-D\nabla^2 -\overline{F}'(u(x,t))\right )G(x,t;x',t') d =\delta(x-x')\delta(t-t'). 
\end{eqnarray}
Assuming the initial condition $\widetilde{U}(x,0)=0$, it follows that 
\begin{equation}
 \widetilde{U}(x,t)=\int_0^t\int_{\R} G(x,t;x',t')\widetilde{\xi}(x',t')dx'dt'
\end{equation}
and $\langle  \widetilde{U}(x,t)\rangle =0$.
Equation (\ref{xi2}) then yields
\begin{eqnarray}
\label{dodo}
\fl  \Delta(x,y,t)=2  \int_0^t  \int_{\R}  \int_{\R}  G(x,t;x',t')  G(y,t;x'',t')   C(u(x',t'),u(x'',t')) dx'\, dx'' dt'.
\end{eqnarray}
Differentiating both sides using the equation for $G$ gives
\begin{eqnarray}
\fl  &\left (\frac{\partial}{\partial t}-D\frac{\partial^2}{\partial x^2}-D\frac{\partial^2}{\partial y^2}-\overline{F}'(u(x,t))-\overline{F}'(u(y,t)) \right )\Delta(x,y,t)=2   C(u(x,t),u(y,t)) .\nonumber \\
\fl
\label{CCC}
\end{eqnarray}
Finally noting that $u-\langle U\rangle =O(\epsilon)$ we have, to leading order in $\epsilon$
\begin{eqnarray}
\fl  &\frac{\partial \Delta}{\partial t}=D\frac{\partial^2\Delta}{\partial x^2}+D\frac{\partial^2\Delta}{\partial y^2} +\overline{F}'(\langle U(x,t)\rangle)+\overline{F}'(\langle U(y,t)\rangle)  +2   C(\langle U(x,t)\rangle,\langle U(y,t)\rangle) .\fl\label{CCC2}
\end{eqnarray}

\subsection{Pair annihilation}

Suppose that when the substrate is active the particles undergo diffusion-limited pair annihilation \cite{Tauber05,Cardy08}:
\[A+A\overset{\gamma}{\rightarrow} \emptyset + \mbox{ diffusion}.\]
One mechanism for generating substrate-dependent pair annihilation could be adsorption under fast dimerization. Let $U$ denote the concentration of monomers and $\widetilde{U}$ the concentration of dimers. Consider the generalized RD system
\numparts
\begin{eqnarray}
\label{swdiff2}
  \frac{\partial U}{\partial t}&=D\frac{\partial^2U}{\partial x^2}-\gamma U^2+\eta \widetilde{U},\\
  \frac{\partial \widetilde{U}}{\partial t}&=D\frac{\partial^2\widetilde{U}}{\partial x^2}+\gamma U^2-\eta \widetilde{U}-\kappa \delta_{n,1}\widetilde{U}
  \end{eqnarray}
\endnumparts
for $N(t)=n$. Here $\gamma$ is the rate of dimerization under the reaction scheme 
\[A+A\Markov{\eta}{\gamma} 2A,\]
and adsorption only occurs for dimers at a rate $\kappa$ when $N(t)=1$. Furthermore, suppose that $\kappa \rightarrow \infty$ (instant adsorption of dimers) so that $\widetilde{U}(t)=0$ whenever $N(t)=1$, and hence $\eta \widetilde{U} \ll \gamma U^2$ when $N(t)=2$. (The latter condition will hold provided that the switching rates $\alpha,\beta$ are sufficiently fast.)
Under these conditions, we can take the effective adsorption rate of $U$ to be $-\gamma U^2$.

In the case of pair annihilation, we have
 $\overline{F}(u)=-\overline{\gamma} u^2$,  $\overline{\gamma} =\rho_1\gamma$, and
 \begin{equation}
 C(u(x),u(x'))=\sigma^2(u(x)u(x'))^2,\quad \sigma^2=\frac{\alpha \beta \overline{\gamma}^2 }{(\alpha+\beta)^3}.
\end{equation}
First, consider spatially uniform solutions for which equations (\ref{pyg2}) and  (\ref{CCC}) become 
\numparts
\begin{eqnarray}
\label{pyg2a}
&\frac{d\langle U\rangle}{d t}=-\overline{\gamma} \langle U\rangle^2 -  \overline{\gamma} \epsilon \Delta(t),\\
 & \frac{d\Delta}{d t}=-4\overline{\gamma} \langle U\rangle \Delta +4 \sigma^2\langle U\rangle^4 .
 \label{pyg2b}
\end{eqnarray}
\endnumparts
 For finite times we can use regular perturbation theory. Substituting the series expansions 
\[\langle U\rangle=u_0+\epsilon u_1+\ldots,\quad  \Delta=\Delta_0+\epsilon \Delta_1+\ldots \]
into equations (\ref{pyg2a}) and (\ref{pyg2b}), and collecting equal powers of $\epsilon$ yields a hierarchy of equations, the first few of which are as follows:
\numparts
\begin{eqnarray}
\frac{du_0}{d t}&=-\overline{\gamma}  u_0^2(t),\quad u_0(0)=a,\\
\label{harry}
\frac{du_1}{d t}&=-2\overline{\gamma}  u_0(t)u_1(t)-  \overline{\gamma} \Delta_0(t),\quad u_1(0)=0,\\
\frac{d\Delta_0}{d t}&=-4\overline{\gamma}  u_0(t) \Delta_0(t) +4 \sigma^2u_0^4(t),\quad \Delta_0(0)=0.
\end{eqnarray}
\endnumparts
The first equation has the explicit solution
\begin{equation}
u_0(t)=\frac{1}{a+\overline{\gamma}  t}. 
\end{equation}
Plugging into the third equation implies that
\begin{eqnarray*}
\frac{d\Delta_0}{d t}&=-\frac{4\overline{\gamma}}{a+\overline{\gamma}  t}\Delta_0(t) +\frac{4 \sigma^2}{(a+\overline{\gamma}  t)^4},
\end{eqnarray*}
which can be solved using an integrating factor to give
\begin{equation} \Delta_0(t)= \frac{ 4\sigma^2t}{ (a+\overline{\gamma}  t)^4}.
\end{equation}
Finally, substituting for $u_0,\Delta_0$ into equation (\ref{harry}) gives
\begin{eqnarray*}
\frac{du_1}{d t}&=-\frac{2\overline{\gamma}}{a+\overline{\gamma}t}u_1(t)-  \frac{4\sigma^2\overline{\gamma}t}{(a+\overline{\gamma}  t)^4},
\end{eqnarray*}
which has the solution
\begin{equation}
\fl u_1(t)= -\frac{4\sigma^2}{ \overline{\gamma}(a+\overline{\gamma}  t)^2}\int_0^{\overline{\gamma}t}\frac{x}{(a+x)^2}dx= -\frac{4\sigma^2}{ \overline{\gamma}(a+\overline{\gamma}  t)^2}\left [\ln (1+\overline{\gamma}t/a) -\frac{\overline{\gamma}t}{a+\overline{\gamma} t}\right ].
\end{equation}
Combining the various results, we obtain finite-$t$ behavior
\numparts
\begin{eqnarray}
\label{ann1}
\fl \langle U(t)\rangle &= \frac{1}{a+\overline{\gamma}  t} - \frac{4\sigma^2\epsilon}{ \overline{\gamma}(a+\overline{\gamma}  t)^2}\left [\ln (1+\overline{\gamma}t/a) -\frac{\overline{\gamma}t}{a+\overline{\gamma} t}\right ]+O(\epsilon^2),\\ 
\fl \Delta(t)&= \frac{ 4\sigma^2t}{ (a+\overline{\gamma}  t)^4}+O(\epsilon).
\end{eqnarray}
\endnumparts

Equation (\ref{ann1}) suggests that the $O(\epsilon)$ correction to the mean field equation due to switching decays faster than the leading order term, so that $u(t)\sim t^{-1}$ independently of the initial density. However, a well-known feature of diffusion-limited pair annihilation is that this form of asymptotic scaling only holds in dimensions higher than two. Indeed, Monte-Carlo simulations of a spatially discretized version of the model (without switching), in which particles randomly hop between neighboring sites on a lattice, reveal more complex behavior that is dependent on the spatial dimension $d$ \cite{Reid03}:
\begin{equation}
\label{eq16:asy}
u \sim t^{-1/2} \ (d=1),\ u\sim t^{-1}\ln t\ (d=2) ,\ u\sim t^{-1} \ (d> 2).
\end{equation}
This is a consequence of the fact that 
the RD equation is a macroscopic equation for the mean particle density $u(x,t)$, which ignores any spatial fluctuations and statistical correlations between particles due to low copy numbers. One way to understand the breakdown of the macroscopic RD equation at $d=2$ is to recall that random walks display certain universal properties. In particular, an unbiased random walk on a $d$-dimensional lattice is recurrent if $d\leq 2$ and transient if $d>2$. In a decay process such as pair annihilation, the surviving particles at large times are separated by large distances. This means that the probability of a pair of diffusing particles coming into close proximity to annihilate each other is strongly dependent on whether diffusion is recurrent or transient. For $d\leq 2$ (recurrent diffusion) a pair of particles find each other with probability 1, even if they are represented by points in a continuum limit. Hence, the effective diffusion-limited reaction rate will be governed by universal features of diffusion. On the other hand, if $d>2$ (transient diffusion) then the probability of point particles meeting vanishes. That is, for any reaction to occur, the particles must have finite size or reaction radius, or be placed on a lattice. The effective reaction rate will then depend on the microscopic details of the short-distance spatial regularization, meaning that a degree of universality is lost. The occurrence of universal behavior at and below some dimension $d_c$, known as the upper critical dimension, is typically indicative of a break down of macroscopic mean field equations due to the effects of statistical fluctuations. In the case of binary reactions such as pair annihilation, $d_c=2$. In classical field theoretic treatments of pair annihilation with diffusion \cite{Tauber05,Cardy08,Atland10,Kamenev14,Tauber14}, the effects of small molecular numbers at each lattice site is dealt with by considering an RD master equation, in which diffusive hopping between neighboring lattice sites is treated as an additional set of single step reactions that supplement the pair-annihilation reaction. Path integral and renormalization group methods can then be used to extract the correct scaling laws.

\section{Path integral for a hybrid RD master equation}

The derivation of the hybrid path integral (\ref{dpi}) in section 3 assumed a spatial discretization scheme in which the number of molecules per lattice site was sufficiently large so that we could define a local concentration in the continuum limit. However, the particular example of pair annihilation considered in section 5.3 suggests that there are situations in which the existence of a local concentration breaks down. Replacing a hybrid RD differential equation by a hybrid RD master equation means that fluctuations due to a switching environment are supplemented by an additional source of fluctuations due to low molecular numbers. This means that one has to specify the $\epsilon$-scaling of both sources of noise when developing a path integral representation in the semi-classical limit. Analogous observations have previously been made within the context of non-spatial models of stochastic ion channels \cite{Keener11} and gene networks \cite{Kepler01,Newby15,Li16}.

\subsection{Hybrid RD master equation}
Returning to the spatial discretization scheme of section 2, we now set $u_{j}=r_{j}/h^d$ at the $j$-th lattice site, where $r_j$ is the number of molecules and $\Omega=h^d$ is the hypervolume of a single discrete cell ( in $d$ spatial dimensions). In order to write down an RD master equation, we decompose the mass action functions $F_n(u)$ into a set of $K$ single-step reactions according to 
\begin{equation}
F_n(u)=\sum_{\mu=1}^KS_{\mu}f_n^{\mu}(u),
\end{equation}
where $S_{\mu}$ is a stoichiometric coefficient and the functions $f_n^{\mu}$ are $n$-dependent propensities. (The extension to more than one chemical species is straightforwardly defined in terms of a corresponding stoichiometric matrix.)
If $r_j$ is sufficiently large, then we can treat $r_i$ as a continuous variable of time and write down the analog of equation (\ref{sh}):
\begin{equation}
\label{eq16:RD2}
\frac{dr_i}{d t}= \sum_{j\in \calJ}\Delta_{ij}r_j +h^d \sum_{\mu=1}^KS_{\mu}f_n^{\mu}(r_i/h^d),\quad i\in \calJ 
\end{equation}
when $N(t)=n$. 
If one views equation (\ref{eq16:RD2}) as the rate equation of a Markov process with discrete variables $\r(t)=\sum_{j} r_{j}(t){\bf e}_{j}$, $(\e_j)_i=\delta_{i,j}$, then the associated master equation for the probability distribution $P_n(\r,t)$, where $n$ is the discrete state of the environment, is
\begin{eqnarray}
\fl \frac{dP_n(\r,t)}{dt}
 &=\frac{D}{h^2}\sum_{\langle j,j'\rangle}[(r_{j}+1)P_n(\r+{\bf e}_{j}-{\bf e}_{j'},t)-r_{j}P_n(\r,t)] \nonumber \\
\fl &  \quad +\frac{D}{h^2} \sum_{\langle j,j'\rangle}[(r_{j'}+1)P_n(\r-{\bf e}_{j}+{\bf e}_{j'},t)-r_{j'}P_n(\r,t)]  \\ 
\fl &\quad +h^d\sum_j \sum_{\mu=1}^K \bigg (f_n^{\mu}([r_j-S_{\mu}]/h^d) P(\r-S_{\mu} {\bf e}_j,t)-f_n^{\mu}(r_j/h^d)P(\r,t)\bigg )\nonumber \\
\fl &\quad +\sum_{m}Q_{nm}P_m(\r,t),\nonumber 
\label{RDme}
\end{eqnarray}
where $\langle j,j'\rangle$ indicates that we are summing over nearest neighbors on the lattice without double counting. 
Note that $\r+{\bf e}_{j}-{\bf e}_{j'}=(\ldots,r_{j}+1,r_{j'}-1,\ldots)$ etc. The first and second lines represent single-step hopping transitions $j\rightarrow j'$ and $j'\rightarrow j$, respectively, while the third and fourth lines represent the set of chemical reactions at each lattice site and changes in environmental state, respectively.

\subsection{Doi-Peliti operators}
The next step is to convert the RD master equation to operator form by introducing Doi-Peliti annihilation and creation operators $a_j,a_j^{\dagger}$ at each lattice site \cite{Doi76,Peliti85,Tauber05,Cardy08}. The corresponding commutation relations for $a_i,a_i^{\dagger}$ are
\begin{equation}
[a_i,a_j^{\dagger}]=\delta_{i,j},\ [a_i,a_j]=[a^{\dagger}_i,a^{\dagger}_j].
\end{equation}
Define the state vectors
\begin{eqnarray}
|\psi_n(t)\rangle 
&=\sum_{\r} P_n(\r,t)|\r\rangle  ,
\end{eqnarray}
where $\sum_{\r}=\prod_{j \in \Z^d}\sum_{r_{j}\geq 0}$ and $|\r\rangle=\prod_j(a_j^{\dagger})^{r_j}|0\rangle$. In addition,
\begin{eqnarray*}
& a_j|\r\rangle =r_j|\ldots,r_{j-1},r_j-1,r_{j+1},\ldots\rangle, \\
& a_j^{\dagger}|\r\rangle =|\ldots,r_{j-1},r_j+1,r_{j+1},\ldots\rangle .
\end{eqnarray*}
Differentiating with respect to time and plugging in the master equation leads to the following operator version of the master equation:
\begin{equation}
\frac{d}{dt}|\psi_n(t)\rangle = \widehat{K}_n|\psi_n(t)\rangle+\sum_{m=0,1}Q_{nm}|\psi_m(t)\rangle ,
\label{opmaster}
\end{equation}
where
\begin{eqnarray}
\fl \widehat{K}_{n}&=-\frac{D}{h^2} \sum_{\langle j,j'\rangle} (a_{j'}^{\dagger}-a_{j}^{\dagger})(a_{j'}-a_{j})  +h^d\sum_j\sum_{\mu=1}^K\gamma_n^{\mu}(a_j/h^d,a_j^{\dagger}/h^d) .
\label{Kn}
\end{eqnarray}
Each local term $\gamma_n^{\mu}$ is a polynomial involving products of annihilation and creation operators scaled by $h^d$.

An important step in the construction of a Doi-Peliti path integral is the use of the coherent-state representation
 \begin{equation}
{ |{\bm \varphi}\rangle = \exp\left (-\frac{1}{2}\sum_j|\varphi_i|^2 \right )\exp\left (\sum_i\varphi_i a_i^{\dagger}\right )|0\rangle,}
 \end{equation}
where $\varphi_i$ is the complex-valued eigenvalue of the annihilation operator $a_i$, with complex conjugate ${\varphi}_i^*$. The coherent states satisfy the completeness relation
\begin{equation}
\label{bdcom2}
\int\prod_i\frac{d\varphi_i d{\varphi}_i^* }{\pi}|{\bm \varphi}\rangle \langle {\bm \varphi}|=1.
\end{equation}
In order to determine the action of $\widehat{K}_{n}$ on $|{\bm \varphi}\rangle$ it is first necessary to normal-order $\widehat{K}_{n}$ by moving all creation operators $a_j^{\dagger}$ to the left of all annihilation operators $a_j$ using the commutation relations. The operator $\widehat{K}_n$ then becomes
\begin{eqnarray}
\widehat{K}_{n}&=-\frac{D}{h^2} \sum_{\langle j,j'\rangle} (a_{j'}^{\dagger}-a_{j}^{\dagger})(a_{j'}-a_{j}) +h^d\sum_j \Gamma_n(a_j/h^d,a_j^{\dagger}/h^d),
\end{eqnarray}
with
\begin{equation}
\Gamma_n(a_j/h^d,a_j^{\dagger}/h^d)=\widehat{\mathcal N}\bigg \{ \sum_{\mu=1}^K\gamma_n^{\mu}(a_j/h^d,a_j^{\dagger}/h^d) \bigg \}
\end{equation}
and $\widehat{\mathcal N}$ the normal-ordering operator. It now follows that
\begin{eqnarray}
\fl \widehat{K}_{n} |{\bm \varphi}\rangle &=\bigg \{-\frac{D}{h^2} \sum_{\langle j,j'\rangle} (\varphi_{j'}^{*}-\varphi_{j}^{*})(\varphi_{j'}-\varphi_{j}) +h^d\sum_j \Gamma_n (\varphi_j/h^d,\varphi_j^*/h^d) \bigg \} |{\bm \varphi}\rangle .\end{eqnarray}
Equation (\ref{opmaster}) can be rewritten in the vector form (\ref{psiH}):
\begin{eqnarray}
\frac{d}{dt}|  {\bm \psi}(t)\rangle&=\widehat{\bf H}|  {\bm \psi}(t)\rangle,\quad |  {\bm \psi}(t)\rangle=(|  {\psi}_1(t)\rangle,\ldots,|  { \psi}_M(t)\rangle)^{\top},
\end{eqnarray} 
where equation (\ref{sh:H}) becomes
\begin{eqnarray}
\widehat{\bf H}&=\mbox{diag}(\widehat{K}_{1} ,\ldots,\widehat{K}_{M})+{\bf Q}.
\end{eqnarray}

For the sake of illustration, consider the example of a single-step reaction based on pair annihilation (see also section 5.3): $A+A\overset{\gamma_n}{\rightarrow} \emptyset$ with an $n$-dependent reaction rate $\gamma_n$. The third line of equation (\ref{RDme}) reduces to the explicit form 
\begin{eqnarray}
 h^{-d}{\gamma_n} \sum_{j} [(r_{j}+2)(r_{j}+1)
P_n(\r+2\e_{j},t) -r_j(r_j-1)P_n(\r,t)].
\end{eqnarray}
(More precisely, equation (\ref{RDme}) assumes that $r_j$ is sufficiently large so that one can drop the constant shifts in $r_j$.) The corresponding operator (\ref{Kn}) is
\begin{eqnarray}
\fl \widehat{K}_{n}&=-\frac{D}{h^2} \sum_{\langle j,j'\rangle} (a_{j'}^{\dagger}-a_{j}^{\dagger})(a_{j'}-a_{j}) + h^{-d}\sum_{j} \left [  a_{j}^2  -a_{j}^{\dagger}a_{j}(a_{j}^{\dagger}a_{j}-1)\right ] .
\end{eqnarray}
Normal ordering the second term on the right-hand side thus gives
\begin{equation}
h^d \Gamma_n (\varphi_j/h^d,\varphi_j^*/h^d)=h^{-d}\gamma_n  \sum_{j} [1-(a_{j}^{\dagger})^2] a_{j}^2.
\end{equation}

\subsection{Path integral}

Following section 3, we divide the time interval $[0,t]$ into $N$ subintervals of size $\Delta t=t/N$ and write
\begin{equation}
|{\bm \psi}(t)\rangle =\e^{\widehat{\bf H} \Delta t}\e^{\widehat{\bf H} \Delta t}\cdots \e^{\widehat{\bf H}\Delta t}\e^{\sum_{j}a_{j}}|{\bm \psi}(0)\rangle.
\end{equation}
We then insert multiple copies of the completeness relations (\ref{comp2}) and (\ref{bdcom2}) for the basis vectors $|s,{\bm \varphi}\rangle=|s\rangle \otimes |{\bm \varphi}\rangle$, where $|s\rangle$ is the spinor (\ref{spinor}), so that
\begin{eqnarray}
& |{\bm \psi}(t)\rangle =\int_{\Omega}ds_0 \cdots \int_{\Omega}ds_N\int \frac{d{\bm \varphi}_0 d{\bm \varphi}_0^* }{\pi}\, \cdots \int \frac{d{\bm \varphi}_N d{\bm \varphi}_N^* }{\pi}\nonumber 
\\
& \times |s_N,{\bm \varphi}_N\rangle \prod_{\ell=0}^{N-1}\langle s_{\ell+1},{\bm \varphi}_{\ell+1}|\e^{\widehat{\bf H} \Delta t}|s_{\ell},{\bm \varphi}_{\ell}\rangle \langle s_{0}, \bm {\varphi}_{0}|{\bm \psi}(0)\rangle .
\label{pipall}
\end{eqnarray}
In the limit $N\rightarrow \infty$ and $\Delta t \rightarrow 0$ with $N\Delta t =t$ fixed, we can make the approximation
\begin{eqnarray}
  \label{st}
\fl  &\langle s_{\ell+1},{\bm \varphi}_{\ell+1}|\e^{\widehat{\bf H} \Delta t}| s_{\ell},{\bm \varphi}_{\ell}\rangle \approx \langle s_{\ell+1}|s_{\ell}\rangle \langle{\bm \varphi}_{\ell+1}| {\bm \varphi}_{\ell} \rangle \bigg (1 + H({\bf c}_{\ell},{\bm \phi}_{\ell},{\bm \varphi}_{\ell},{\bm \varphi}^*_{\ell})\Delta t \bigg ),
\end{eqnarray}
where
\begin{eqnarray}
\label{odg2}
\fl H({\bf c},{\bm \phi},{\bm \varphi},{\bm \varphi}^*)&\equiv \langle s|\widehat{\bf H}|s\rangle=-\frac{D}{h^2} \sum_{\langle j,j'\rangle} (\varphi_{j'}^{*}-\varphi_{j}^{*})(\varphi_{j'}-\varphi_{j}) \\
 \fl &\quad -h^d\sum_{m=1}^M c_m \Gamma_m (\varphi_j/h_d,\varphi_j^*/h_d) + \sum_{\underset{n\neq m}{n,m=1}}^MT_{nm}\left [\e^{i(\phi_m-\phi_n)/2}-1\right ]c_m.
 \nonumber
\end{eqnarray}
The inner product $\langle s_{\ell+1}|s_{\ell}\rangle$ satisfies equation (\ref{sl}). 
In addition, using standard properties of coherent states, 
\begin{eqnarray}
\fl &\langle {\bm \varphi}_{\ell+1}|{\bm \varphi}_{\ell}\rangle= \exp\left (\frac{1}{2}\sum_j|\varphi_{\ell+1,i}|^2-\frac{1}{2}\sum_j|\varphi_{\ell,i}|^2 \right ) \exp\left (-\sum_i\varphi^*_{\ell+1,i }(\varphi_{\ell+1,i}-\varphi_{\ell,i})\right ) .\nonumber \\
\fl
\end{eqnarray}
Finally, substituting the expression for the small-time propagator back into equation (\ref{pipall}) yields 
\begin{eqnarray*}
\fl  & |{\bm \psi}(t)\rangle =\int_{\Omega}ds_0 \cdots \int_{\Omega}ds_N\int \frac{d{\bm \varphi}_0 d{\bm \varphi}_0^* }{\pi}\, \cdots \int \frac{d{\bm \varphi}_N d{\bm \varphi}_N^* }{\pi}  |s_N,{\bm \varphi}_N\rangle \langle s_{0}, \bm {\varphi}_{0}| {\bm \psi}(0)\rangle \\ 
\fl &\ \times\prod_{\ell=0}^{N-1} \exp\bigg (\bigg[H({\bf c}_{\ell},{\bm \phi}_{\ell},{\bm \varphi}_{\ell},{\bm \varphi}^*_{\ell})-\frac{i}{2}\sum_{m=1}^M\frac{d\phi_{\ell,m}}{dt}c_{\ell,m}+i{\bm \varphi}^*_{\ell}\cdot \frac{d{\bm \varphi}_{\ell}}{dt}\bigg ]\Delta t\bigg )  . \nonumber
\end{eqnarray*}
We now take the limit $N\rightarrow \infty,\Delta t\rightarrow 0$ with $N\Delta t=t$ fixed, ${\bm \varphi}_{\ell}={\bm \varphi}({\ell}\Delta t)$ and ${\bm \varphi}_{\ell}^*={\bm \varphi}^*({\ell}\Delta t)$. After Wick ordering, ${\bm \varphi}^*\rightarrow -i{\bm \varphi}^*$ and integrating by parts the term involving $d\phi/dt$, we obtain the following functional path integral:
\begin{eqnarray}
 \fl & |{\bm \psi}(t)\rangle \sim  \int  {\mathcal D}[z]  {\mathcal D}[\phi]{\mathcal D}[{\bm \varphi}]{\mathcal D}[{\bm \varphi}^*] \varphi_j(t)
  \exp\left (-\int_{0}^{t}\left [{\bm \varphi}^*\cdot \frac{d{\bm \varphi}}{d\tau}-i\phi\frac{dz}{d\tau} -{\mathcal H}\right ]d\tau\right ), 
\label{pieff2}
\end{eqnarray}
where ${\mathcal H}$ is the effective Hamiltonian (\ref{odg2}). We can now take the continuum limit by letting the lattice spacing $h\rightarrow 0$ such that $\sum_{j} h^d \rightarrow \int d\x$ and rescaling the fields according to
\begin{eqnarray}
h^{-d}\varphi_{j}(t) \rightarrow \varphi(\x,t),\ \varphi_{j}^*(t)\rightarrow \widetilde{\varphi}(\x,t).
\label{sys}
\end{eqnarray}
The result is 
\begin{eqnarray}
\label{rdpath}
&|{\bm \psi}(t)\rangle \sim \int {\mathcal  D}[{\bf c}]{\mathcal  D}[{\bm \phi }]{\mathcal  D}[\varphi ]\, {\mathcal  D}[\widetilde{\varphi} ]\, \e^{-  S[{\bf c},{\bm \phi},\varphi,\widetilde{\varphi}]},
\end{eqnarray}
with the action functional
\begin{eqnarray}
\fl S&=\int_0^t \bigg \{\int_{\R^d} \left [\widetilde{\varphi}\left (\frac{\partial \varphi}{\partial \tau}-D\nabla^2\varphi \right )\right ]d\x -\sum_{m=1}^M c_m\int_{\R^d} \Gamma_m( {\varphi,\widetilde{\varphi}}) d\x  \nonumber  \\
\fl & \qquad \qquad -\frac{i}{2} \sum_{m=1}^M\phi_m\frac{dc_m}{d\tau}  -\sum_{\underset{n\neq m}{n,m=1}}^MT_{nm}\left [\e^{i(\phi_m-\phi_n)/2}-1\right ]c_m \bigg\}d\tau .
\end{eqnarray}

\subsection{Semi-classical limit}

One of the major differences between the hybrid PDE model of an RD process considered in sections 2-5 and the hybrid RD master equation is that the latter has two sources of noise: fluctuations due to environmental switching and fluctuations due to low molecular numbers. This implies that we have to introduce separate scalings in the weak-noise limit, analogous to a previous study of a non-spatial gene network \cite{Li16}. The fast switching or adiabatic limit is implemented as in section 4 by taking
${\bf T}\rightarrow {\bf T}/\epsilon$. On the other hand the limit of weak molecular noise requires a system size scaling as suggested by equation (\ref{sys}). Therefore, we take
$\varphi \rightarrow \varphi/\epsilon^{\nu}$, $\nu >0$, and rescale the various reaction rates such that $\Gamma_n(\varphi/\epsilon^{\nu},\widetilde{\varphi}) \rightarrow \Gamma_n(\varphi,\widetilde{\varphi})/\epsilon^{\nu}$. The path integral (\ref{rdpath}) becomes
\begin{eqnarray}
\label{rdpath2}
&|{\bm \psi}(t)\rangle \sim \int {\mathcal  D}[{\bf c}]{\mathcal  D}[{\bm \phi }]{\mathcal  D}[\varphi ]\, {\mathcal  D}[\widetilde{\varphi} ]\,  \e^{-  S_1[{\bf c},\varphi,\widetilde{\varphi}]/\epsilon^{\nu}- S_2[{\bf c},{\bm \phi}]/\epsilon},
\end{eqnarray}
where
\numparts
\begin{eqnarray}
\fl S_1&=\int_0^t \bigg \{\int_{\R^d} \left [\widetilde{\varphi}\left (\frac{\partial \varphi}{\partial \tau}-D\nabla^2\varphi \right )\right ]d\x -\sum_{m=1}^M c_m\int_{\R^d} \Gamma_m( {\varphi,\widetilde{\varphi}}) d\x \bigg\}d\tau , \\
\fl S_2 &=\int_0^t \bigg \{ -\frac{i}{2} \sum_{m=1}^M\phi_m\frac{dc_m}{d\tau}  -\sum_{\underset{n\neq m}{n,m=1}}^MT_{nm}\left [\e^{i(\phi_m-\phi_n)/2}-1\right ]c_m \bigg\}d\tau .
\end{eqnarray}
\endnumparts

Elimination of the auxiliary variables now depends on the value of $\nu$. If $\nu >1$ then environmental switching is faster than the chemical reactions, and the action $S_2$ dominates in the limit $\epsilon \rightarrow 0$. One thus finds that $c_m \rightarrow \rho_m$, where ${\bm \rho}$ is the stationary distribution of the Markov chain with generator ${\bf Q}$. The semi-classical path integral becomes
\begin{eqnarray}
 & |{\bm \psi}(t)\rangle  \sim \int   {\mathcal D}[\varphi]{\mathcal D}[\widetilde{\varphi}]    \e^{-  \overline{S}_1[\varphi,\widetilde{\varphi}]/\epsilon^{\nu}} ,
 \label{path0}
\end{eqnarray}
where 
\begin{eqnarray}
 \overline{S}_1&=\int_0^t \bigg \{\int_{\R^d} \left [\widetilde{\varphi}\left (\frac{\partial \varphi}{\partial \tau}-D\nabla^2\varphi \right )\right ]d\x + \int_{\R^d}\overline{\Gamma}( {\varphi,\widetilde{\varphi}}) d\x \bigg\}d\tau ,\end{eqnarray}
and $\overline{\Gamma}=\sum_{m=1}^M\rho_m \Gamma_m$. That is, we obtain the path integral for the RD master equation in which reaction rates are averaged with respect to the distribution ${\bm \rho}$. This is the mean-field version of the switching system. On the other hand, if $\nu<1$ then the chemical reactions are faster than environmental switching, and the action $S_1$ dominates in the limit $\epsilon \rightarrow 0$. In order to interpret the semi-classical limit, we would need to make further assumptions about the chemical kinetics in a fixed environment. Therefore, we will focus on the most difficult case, namely $\nu=1$, whereby the rates of environmental switching and chemical reactions are comparable. One useful feature of representing the environmental states in terms of spinors is that the elimination of the auxiliary variables proceeds straightforwardly along identical lines to section 4. Therefore, we can simply write down the resulting path-integral in the semi-classical limit:
\begin{eqnarray}
 & |{\bm \psi}(t)\rangle  \sim \int    {\mathcal D}[\varphi]{\mathcal D}[\widetilde{\varphi}]    \e^{-  S[\varphi,\widetilde{\varphi}]/\epsilon} ,
 \label{path1}
\end{eqnarray}
where
\begin{eqnarray}
  S[\varphi,\widetilde{\varphi}]&=\int_{0}^{t} \bigg \{\int_{\R^d}  \left [\widetilde{\varphi}\left (\frac{\partial \varphi}{\partial \tau}-D\nabla^2\varphi \right )\right ]d\x -\Lambda[\varphi,\widetilde{\varphi}]\bigg\}d\tau,
\end{eqnarray}
and $\Lambda$ is the Perron or principal eigenvalue of the functional eigenvalue equation
\begin{eqnarray}
 \fl &\sum_{m=1}^M\left \{\delta_{m,n}  \int_{\R^d} \Gamma_n[\varphi,\widetilde{\varphi}]d\x+Q_{nm}\right \}R_{m}[\varphi,\widetilde{\varphi}] = {\Lambda}[\varphi,\widetilde{\varphi}]R_{n}[\varphi,\widetilde{\varphi}].
 \label{pathmaster}
\end{eqnarray}
The path-integral (\ref{path1}) provides a framework for analyzing hybrid RD equations in the weak-noise limit, based on a combination of large deviation theory and Gaussian approximations as outlined in section 4 for the hybrid PDE. For example, least action paths satisfy the functional Hamilton-Jacobi equation
\begin{equation}
H[\varphi,\delta_{\varphi}\Phi]=0,\quad \delta_{\varphi}\Phi=\frac{\delta \Phi}{\delta \varphi(\x)},
\label{HJ2}
\end{equation}
with
\begin{equation}
\label{ham2}
H[\varphi,\widetilde{\varphi}]=D\int_{\R^d} \widetilde{\varphi}(\x) \nabla^2\varphi(\x)d\x+\Lambda[\varphi,\widetilde{\varphi}].
\end{equation}

\section{Discussion}

In this paper we constructed path integrals for stochastic hybrid RD processes whose reaction terms depended on the discrete state of a randomly switching environment. The construction was based on the application of operator methods to a spatially discretized version of a given RD system, in which the environmental states were represented in terms of spinors. The spinor representation facilitated the elimination of the auxiliary path integral variables $c_m(t)$, $m=1,\ldots,M$, and their conjugates in the fast-switching or adiabatic limit. The variable $c_m(t)$ determined the effective probability that a sample path was exposed to the environmental state $m$ at time $t$ with $\sum_{m=1}^Mc_m(t)=1$. The elimination of the auxiliary variables generated an effective action whose Hamiltonian was the sum of a diffusion term and the Perron or principal eigenvalue of a functional linear operator involving the reaction terms and the matrix generator of the switching process. The reduced path integral was then used to derive a functional Hamilton-Jacobi equation for least action paths and to obtain a Gaussian noise approximation of the stochastic hybrid RD system in the adiabatic limit.
Finally, the path integral construction was generalized to the case of a RD master equation, which combined stochastic environmental switching with fluctuations due to molecular noise. Although we restricted our treatment to a single molecular species for notational convenience, it would be straightforward to consider more general multi-species RD systems.

There are number of outstanding issues that warrant further consideration. First, finding non-trivial solutions of the functional Hamilton-Jacobi equation (\ref{HJ}) or (\ref{HJ2}). These could represent least-action paths of escape from a metastable state in the weak noise limit, in an analogous fashion to finite-dimensional hybrid systems \cite{Bressloff13a,Bressloff14,Bressloff15}. A crucial step in this analysis would be to solve the corresponding functional linear equation (\ref{eigR}) or (\ref{pathmaster}) for the Perron eigenvalue. A second issue concerns the development of more systematic diagrammatic perturbation methods for calculating corrections to the mean-field RD equations in the presence of a switching environment and possibly molecular noise. 

Finally, it would be interesting to go beyond the adiabatic limit by working with the full path integral representations (\ref{pieff}) or (\ref{rdpath}), in which the auxiliary variables are included. There are several studies of non-spatial models of gene expression that have explored stochastic dynamics in the non-adiabatic regime \cite{Sasai03,Zhang13,Chen15,Bhatt20}. The basic idea of these studies is to carry out a ``small momentum expansion'' with respect to the conjugate variables of the physical degrees of freedom and the auxiliary coordinates $c_m(t)$. The resulting Gaussian approximation of the path integral represents an SDE in an extended phase space that includes the auxiliary variables. For a low-dimensional system one typically finds that the dynamics is driven by a combination of gradient flow along a non-equilibrium potential landscape, which is related to the stationary probability distribution of the multivariate Fokker-Planck equation, and vortex dynamics associated with a non-zero probability flux. However, certain care has to be taken in interpreting these results, since Gaussian noise does not preserve the normalization condition $\sum_{m=1}^M c_m(t)=1$ and the precise physical meaning of the stochastic dynamics in the extended phase space is unclear. Such features become even more problematic in the case of spatial models.

\end{document}